\documentclass[12pt,preprint]{aastex}

\shorttitle{Dust Attenuation in Late-Type Galaxies. I.}
\shortauthors{Pierini et al.}

\begin{document}

\title{Dust Attenuation in Late-Type Galaxies. I. \\
    Effects on Bulge and Disk Components}

\author{D. Pierini}
\affil{Max-Planck-Institut f\"ur extraterrestrische Physik, Postfach 1312,
    D-85741 Garching, Germany}
\email{dpierini@mpe.mpg.de}

\author{K. D. Gordon}
\affil{Steward Observatory, University of Arizona, Tucson, AZ 85721}
\email{kgordon@as.arizona.edu}

\author{A. N. Witt}
\affil{Ritter Astrophysical Research Center, The University of Toledo,
    Toledo, OH 43606}
\email{awitt@dusty.astro.utoledo.edu}

\and

\author{G. J. Madsen}
\affil{Department of Astronomy, University of Wisconsin, Madison, WI 53706}
\email{madsen@astro.wisc.edu}


\begin{abstract}
We present results of new Monte Carlo calculations
made with the DIRTY code of radiative transfer
of stellar and scattered radiation
for a dusty giant late-type galaxy like the Milky Way,
which illustrate the effect of the attenuation of stellar light
by internal dust on the integrated photometry
of the individual bulge and disk components.
Here we focus on the behavior of the attenuation function,
the color excess, and the fraction of light scattered or directly transmitted
towards the outside observer as a function of the total amount of dust
and the inclination of the galaxy, and the structure
of the dusty interstellar medium (ISM) of the disk.
We confirm that dust attenuation produces qualitatively and quantitatively
different effects on the integrated photometry of bulge and disk,
whatever the wavelength.
In addition, we find that the structure of the dusty ISM affects
more sensitively the observed magnitudes than the observed colors
of both bulge and disk.
Finally, we show that the contribution of the scattered radiation
to the total monochromatic light received by the outside observer
is significant, particularly at UV wavelengths,
even for a two-phase, clumpy, dusty ISM.
Thus understanding dust scattering properties is fundamental
for the interpretation of extragalactic observations in the rest-frame UV.
\end{abstract}

\keywords{galaxies: spiral --- dust, extinction --- radiative transfer --- methods: numerical}

\section{Introduction}

Dust plagues the recovery from observations
of the intrinsic (i.e., unaffected by the dust) total and two-dimensional
(i.e., 2-D) photometric parameters of the dusty stellar systems.
Its effects vary as a function of the wavelength
of the photons that are probed by a given broad-band filter
(see Calzetti 2001 for a recent review).

Radiative transfer calculations for simplistic dust/stars configurations,
simulating dusty late-type galaxies, offer equivocal results
for the attenuation by internal dust of the stellar light
produced in these systems (Disney, Davies, \& Phillips 1989).
They also frequently do not describe the observations
over a large spectral range (Kuchinski et al. 1998).
Furthermore, {\it realistic radiative transfer simulations have to include
scattering from dust grains} (Bruzual, Magris, \& Calvet 1988;
Witt, Thronson, \& Capuano 1992; Emsellem 1995),
a phenomenon that is not negligible given the high albedo
observed e.g. for Galactic dust (Gordon, Calzetti, \& Witt 1997; Gordon 2004).
To make computations even more difficult,
{\it dust scattering is not isotropic}: the scattering angle is smaller
the shorter the wavelength of a photon,
as described by the scattering phase function asymmetry parameter
(e.g. Draine 2003; see also Gordon 2004 for a recent review).
The magnitude of the errors introduced by simplification
of the dust/stars configuration, as well as by either neglecting
or approximating dust scattering, depends on the inclination of the galaxy
(Baes \& Dejonghe 2001).

In addition, {\it the dusty interstellar medium (ISM)
present in the disk of a late-type galaxy is observed to be patchy}
(e.g. Rix \& Rieke 1993; Quillen et al. 1995; White, Keel, \& Conselice 2000;
White \& Keel 2001).
Hence, strongly reddened starlight may never dominate
the UV/optical/near-IR radiation that we see.
The local structure (or clumpiness) of the dusty ISM
(Natta \& Panagia 1984; Hobson \& Schueuer 1993; Witt \& Gordon 1996, 2000;
Bianchi, Ferrara, \& Giovanardi 1996; Berlind et al. 1997;
Kuchinski et al. 1998; Bianchi et al. 2000)
and differences in the dust/stars configuration
as a function of the ages of the individual stellar populations
(Silva et al. 1998; Bianchi et al. 2000; Charlot \& Fall 2000;
Popescu et al. 2000; Matthews \& Wood 2001; Tuffs et al. 2004)
affect the main result of the radiative transfer calculations,
i.e., the attenuation function\footnote{{\it The extinction curve describes
the combined absorption and out-of-the beam scattering properties
of a mixture of dust grains of given size distribution and chemical composition
in a screen geometry as a function of wavelength; the attenuation function
is the combination of the extinction curve with the geometry of a dusty
stellar system, in which a substantial fraction of the scattered light
is returned to the line of sight}.}.
As a consequence they affect the estimate
of the total face-on\footnote{The term face-on is used for a disk galaxy
seen at an inclination equal to 0 degrees.} extinction optical depth
along the line of sight through the center of the galaxy
e.g. in the V band (0.55 $\rm \mu m$).
This quantity is referred to as the disk opacity,
and offers some measure of the central dust column density
and, thus, of the total amount of dust present in the disk
(e.g. Misiriotis \& Bianchi 2002).
The uncertain parameterization of the dust clumps for the Milky Way
(cf. Kuchinski et al. 1998 and Bianchi et al. 2000)
and other late-type galaxies leads to maximal uncertainties
on the amount of dust present in a disk galaxy of the order of 40 per cent
(Misiriotis \& Bianchi 2002).
We note that the uncertainties in the dust masses of nearby galaxies
determined from analyzing the far-IR/submm emission from dust
(e.g. Dupac et al. 2003) are at least comparable to those
affecting analogous estimates from extinction modeling alone
(see e.g. Domingue et al. 1998).

Radiative transfer models for realistic dust/stars configurations
simulating late-type galaxies, including multiple scattering,
have been computed either via analytical approximations
(e.g. Byun, Freeman, \& Kylafis 1994; Silva et al. 1998; Xilouris et al. 1999;
Baes \& Dejonghe 2001; Tuffs et al. 2004) or via Monte Carlo techniques
(e.g. de Jong 1996; Wood 1997; Kuchinski et al. 1998; Ferrara et al. 1999;
Bianchi et al. 2000; Matthews \& Wood 2001).
In particular, Kuchinski et al. (1998) made a quantitative investigation
of the opacity of disks using BVRIJHK photometry of 15 highly inclined
Sab--Sc galaxies and the then new DIRTY radiative transfer model
for a bulge$+$disk galaxy (Gordon et al. 2001).
This set of DIRTY models included not only anisotropic multiple scattering
but also two distinctly different, local distributions for Galactic-type dust,
assumed to be present in the disks of all giant late-type galaxies.
Kuchinski et al. found that the maximum optical and near-IR color excesses
observed in near-edge-on\footnote{The term edge-on is used
for a disk galaxy seen at an inclination equal to 90 degrees.} dust lanes
imply total central face-on extinction optical depths of 0.5--2.0
in the V band, consistent with previous results.

Here we adopt the same family of models used by Kuchinski et al.
but with new parameters, as described in Section 2.
We focus on the interpretation of the behavior of the attenuation function,
the color excess, and the fraction of light scattered or directly transmitted
towards the outside observer for the individual bulge and disk components
of a late-type galaxy like our own, as a function of the total amount
of dust and the inclination of the galaxy, and the structure of the dusty ISM
of the disk.
In particular, the present models describe the attenuation
by internal dust at far-UV wavelengths accessible through {\it GALEX}
(Milliard et al. 2001).
Preliminary results have been presented by Pierini, Gordon, \& Witt (2003).

This study not only extends the analysis of Kuchinski et al. (1998),
but also produces necessary, complementary results to those
of Ferrara et al. (1999), Bianchi et al. (2000), and Matthews \& Woods (2001),
which are based on analogous Monte Carlo radiative transfer simulations
for disk galaxies.
In fact, Ferrara et al. (1999) explored the dependence
of the attenuation function on the Hubble type and the inclination
of a disk galaxy, as well as on the total amount of dust,
the dust/stars configuration, and the extinction properties,
for a large wavelength range (14 wavelengths from 0.125 to 2.158 $\rm \mu m$),
but only for a homogeneous dust distribution.
Conversely, Bianchi et al. (2000) investigated the effects of clumping
and the presence of stars embedded in dust clumps
on the total/surface photometry of disks, but only for the V band.
In addition, they assumed a different parameterization of the clumpiness
of the ISM and a different spatial distribution of the clumps
from those assumed by Kuchinski et al. (1998)
or Matthews \& Woods (2001).
In fact, Bianchi et al. (2000) derived the radial distribution
of the dust clumps of their models from the CO observations
of the first Galactic quadrant described by Clemens, Saunders, \& Scoville
(1988), showing the existence of a molecular ring, and assumed all the clumps
to have the same optical depth.
Conversely, clumps have a doubly exponential distribution
and decreasing extinction optical depths as a function
of the galactocentric distance in the models of Kuchinski et al. (1998)
and Matthews \& Woods (2001).
On the other hand, Matthews \& Woods considered, in addition to
the diffuse dust distribution, a two-phase, clumpy dust distribution
where the clumps have a smaller filling factor and a lower density contrast
than those assumed here.
They also made computations for B, R, H, and K bands
and for edge-on disks only.

Finally, this study complements also the analytical description
of radiative transfer presented by Tuffs et al. (2004).
In fact, these authors describe the dependence of the attenuation function
on the Hubble type, the inclination, and the total dust mass
of a late-type galaxy, for 12 wavelengths
ranging from 0.091 to 2.2 $\rm \mu m$.
They consider a dust-less stellar bulge, a disk of old stars
with associated diffuse dust plus a thin disk of young stars
with associated diffuse dust, and a clumpy dust component
associated with star-forming regions in the thin disk
(e.g. Silva et al. 1998)\footnote{Optical opacity measures
in overlapping galaxies suggest that dust is concentrated
in star-forming regions in some disk galaxies (Keel \& White 2001).}.
Their description of the clumpiness approximates the star-forming regions
with porous spheres, where the local absorption of photons
is given by geometry (the solid angle of a parent cloud subtended
at an offspring star of given age), since the optical depths
of the individual clouds are assumed to be extremely large at all wavelengths.
Furthermore, Tuffs et al. (2004) adopt values of the albedo
and, especially, the scattering phase function asymmetry parameter
for Milky Way-type dust that are sensitively different
from those adopted here (Witt \& Gordon 200).

\section{The radiative transfer models}

For the calculations presented in this paper
we employed the DIRTY radiative transfer code (Gordon et al. 2001).
The code applies to any arbitrary spatial distribution of stars and dust.
Here we consider its application to the bulge and disk components
of late-type galaxies separately.

With Monte Carlo techniques (Witt 1977; Audic \& Frisch 1993),
photons are followed through a dust distribution
and their interaction with the dust is parameterized by the optical depth,
the albedo, and the scattering phase function asymmetry parameter,
as defined in e.g. Whittet (2003).
{\it The optical depth determines the probability distribution
for where a photon of given wavelength interacts;
the albedo gives the probability that the photon is scattered
from a dust grain; the scattering phase function
gives the probability distribution for the angle at which the photon scatters}.

For this application, the DIRTY code considers a dust-less stellar bulge,
penetrated by a dusty, stellar disk, as customary (see Ferrara et al. 1999
and Tuffs et al. 2004).
The stellar bulge is modeled as a sphere with a stellar luminosity density
decreasing exponentially with radial distance.
This is at variance with the models of both Ferrara et al. (1999)
and Tuffs et al. (2004), who considered bulges
with $r^{-1/4}$ radial profiles (de Vaucouleurs 1948).
De Vaucouleurs' bulges are more typical of early-type spiral galaxies
(Andredakis, Peletier, \& Balcells 1995; Graham 2001;
Hunt, Pierini, \& Giovanardi 2004).
The outer radius of the bulge model is assumed to be 4 kpc,
with an exponential scale length of 1 kpc, independent of the wavelength
of the tracing photons.
The attenuation effects seen in the bulge are caused
by a doubly exponential dust disk, which extends from the center of the bulge
to the outer edge of the galactic disk component and is described later on.

The stellar disk is modeled as a doubly exponential distribution of sources
with a disk radius of 12 kpc and a scale length of 3 kpc,
independent of the wavelength of the tracing photons
(see Tuffs et al. 2004 for a different description).
Its total height is 2.1 kpc (measured from the galactic plane)
but its scale height is assumed to increase with increasing wavelength
of the tracing photons (see Mihalas \& Binney 1981), as summarized in Table 1.
This variation in the stellar scale height is necessary
to account for the different types of stars that dominate the flux of a galaxy
at different wavelengths: the youngest stars, emitting mostly
at far-UV wavelengths, are more embedded in the dusty medium
than the older ones, emitting mostly in the optical/near-IR.
Other models of radiative transfer deal with this aspect in different ways
(see Charlot \& Fall 2000; Popescu et al. 2000; Tuffs et al. 2004).
We note that the dust-to-star scale-height ratio has a larger impact
on the attenuation function of a disk than the dust-to-star scale-length ratio
(Ferrara et al. 1999).

Dust is present only in a doubly exponential disk with a full radius of 12 kpc,
a scale length of 3 kpc, a scale height of 110 pc,
and a total height of 2.1 kpc (measured from the galactic plane),
that pervades both the stellar disk and the stellar bulge.
It is assumed to consist of dust similar to that found
in the average diffuse ISM of the Milky Way (MW) (Witt \& Gordon 2000;
see Valencic, Clayton, \& Gordon 2004 for a recent study),
its extinction characteristics also summarized in Table 1.
It is locally distributed either in a homogeneous medium
or in a two-phase clumpy one, following the algorithm of Witt \& Gordon (1996).

In the latter case, {\it the clump production process is stochastic
with respect to the spatial coordinates of individual clumps
and the smooth 3-D distribution of the bluest stars}.
In particular, the volume is divided into cubic cells with sides
of 44 pc, the typical diameter of a Galactic giant molecular cloud
(see Blitz 1991), each individual cell being assigned randomly
to either a high-density state or a low-density state by a Monte Carlo process.
The high-density-to-low-density medium contrast is set equal to 100.
The filling factor of the clumps determines the statistical frequency
of the bins in the two states; it is set equal to 0.15 all over the dust disk.
Since the smooth medium has an exponential distribution,
clumps close to the galactic center have a higher extinction optical depth
than those in the outskirts of the dust disk.
In addition, the filling factor of the clumps sets the likelihood that
spatially adjoining cells are occupied by clumps, which leads to the appearence
of complex cloud structures composed of one or more connected clumps.
Two clumps are considered connected when they share one entire side.
The combination of the percolation approach (see e.g. Stauffer 1985)
of generating clouds with the filling factor selected does lead
to a power-law distribution of cloud sizes, as discussed
by Witt \& Gordon (1996)\footnote{In the spherical, two-phase clumpy
models of Witt \& Gordon (1996), the density of a unit cubic clump is identical
throughout a given model, so that the size spectrum is equivalent
to the mass spectrum of the clouds produced. In this case, the synthetic
cloud size-spectrum is very similar to the mass spectrum observed
for interstellar clouds (Dickey \& Garwood 1989). Furthermore,
the fractal dimension of self-similar structures present
in the simulated structured ISM is similar to that observed
for interstellar cloud structures (Vogelaar, Wakker, \& Schwarz 1991).}.

Models are parameterized according to the value of $\tau_V$,
{\it the central face-on extinction optical-depth from the surface
to the mid-plane of the disk in the V band}.
Hence the opacity of the disk (or the total central column density
of the dust) is equal to $2 \times \tau_V$.
For models with the two-phase clumpy dust distribution,
$\tau_V$ is formally somewhat ill-defined (see Witt \& Gordon 1996).
For the two-phase clumpy models, $\tau_V$ is the value
of the face-on extinction optical depth from the surface to the center
of the disk in the V band for the homogeneous model with the same
total dust mass.
Homogeneous and two-phase clumpy models
are computed for $\tau_V = 0.25$, 0.5, 1, 2, 4, and 8.

The large-scale distribution of the dust within the disk
coupled to the parameterization of its clumpiness on the small scale
makes the disk models more optically thick in their inner regions
than at their outskirts (see Kuchinski et al. 1998), whatever the wavelength.
In fact, any photon traveling in a given direction through the dust disk
will face at any step scattering and/or absorption
according to the {\it local} value of the optical depth,
which changes along both the radial and vertical coordinates
as previously described.
For an evaluation of the effects on the attenuation function
for a bulge$+$disk galaxy caused by a different dust/stars configuration
or a different characterization of the clumps, we refer the reader
to Ferrara et al. (1999), Bianchi et al. (2000), and Tuffs et al. (2004).

The dusty bulge$+$disk system is projected onto the plane of the sky,
10 different inclinations (0, 18, 36, 50, 65, 70, 75, 80, 85, and 90 degrees)
being considered.
In addition, 25 wavelengths (from 0.1 to $\rm 3~\mu m$) are included,
with a decimal logarithmic spacing that puts special care
to sampling finely the far-UV domain accessible through {\it GALEX}
(Milliard et al. 2001).
For the wide spectral range considered here,
the scale height of the stellar disk ranges from 60 to 375 pc (see Table 1).

One part of the solution of the radiative transfer model
is the {\it total attenuation optical depth}
at a given wavelength $\lambda$, which is defined as
\begin{equation}
 \tau_{att,~\lambda} = -ln[f_{\lambda}(esc)],
\end{equation}
where $f_{\lambda}(esc)$ is the {\it total} fraction of light
at this wavelength escaping from the system, either directly
or after scattering by dust, in the direction to the outside observer.
$\tau_{att,~\lambda}$ measures the attenuation occurring in the system,
as seen by an observer from a given direction,
and, thus, is {\it an average over many lines of sight within the system}.
Since our model geometry is cylindrically symmetric, $\tau_{att,~\lambda}$
will depend on the inclination $i$ of the galaxy.
We may express the total attenuation as a function of $\lambda$ in magnitudes
(a way preferred by observational astronomers) as follows:
\begin{equation}
 A_\lambda = -2.5~log[f_{\lambda}(esc)] = 1.086 \tau_{att,~\lambda}.
\end{equation}

Given two pass-band filters of effective wavelengths
$\lambda_1$ and $\lambda_2$, we define the color excess
$E(\lambda_1 - \lambda_2)$ as the difference between the values of
the attenuation function (in magnitudes) at $\lambda_1$ and $\lambda_2$,
respectively.

Here we stress that, as a result of the radiative transfer of photons
with different wavelengths through a realistic dusty medium,
the attenuation function $A_\lambda$ (or $\tau_{att,~\lambda}$)
will be different from the extinction curve
assumed for the individual dust grains.
The reason is twofold.
First, radiative transfer determines both absorption and scattering,
for a given extinction curve (see e.g. Gordon et al. 2003),
taking into account the relative distribution of stars and dust
within the system and with respect to the outside observer,
and the structure of the dusty medium (see Witt \& Gordon 1996).
Second, the transfer of radiation is investigated throughout the whole system,
so that even lines of sight different from the observer's one
may contribute to the observed radiation.
For a more extensive explanation, we refer the reader to Calzetti (2001).

\section{Results}

In Sect. 3.1, we investigate how the attenuation by internal dust
of photons of a given wavelength, produced by stars either in the bulge
or in the disk and received by the outside observer,
depends on both the opacity (or total central dust column density)
of the disk (given by $2 \times \tau_V$) and the structure
(two-phase clumpy vs. homogeneous) of the dusty ISM of the disk.
To this purpose, we choose one specific value of the inclination $i$
of the bulge$+$disk galaxy (i.e., $i \rm = 0^o$) and let $\tau_V$ increase
from 0.25 to 8.

Conversely, in Sect. 3.2, we show how much the dust attenuation
for either bulge or disk, and, thus, the observed photometry
of either component, depend on the view angle of the galaxy
and the structure of the dusty ISM of the disk,
once the total amount of dust of the system
and the extinction law of the mixture of dust grains are fixed.
We will discuss models corresponding to $\tau_V = 1$ and $i$ equal to 0, 70
or 90 degrees, since the variation of dust attenuation as a function of $i$
is rather slow between $\rm 0^o$ and $\rm 70^o$
but very fast for $i \rm \ge 70^o$ (see e.g. Ferrara et al. 1999
and Masters, Giovanelli, \& Haynes 2003).

Two different effects are expected to play a role in shaping $A_\lambda$,
when the inclination of the galaxy increases.
One effect is the ``geometrical'' increase of the line-of-sight optical depth
that a photon produced either in the bulge or in the disk faces
once it is inside the dust disk on its journey towards the outside observer
(see e.g. Kylafis \& Xilouris 1996 for the case of edge-on exponential disks).
The other effect is the variation of the fraction of bulge/disk light
which is not affected by the dust disk owing to projection effects.
This variation is different for the bulge and disk structures.
The balance between these two competing effects depends on the wavelength,
since both the extinction curve and the disk stellar distribution
(at least in our models) depend on $\lambda$ (see Sect. 2 and Table 1).

There is no need to assign any spectral energy distribution
to either bulge or disk in order to determine the corresponding
attenuation function, since for either component $A_\lambda$
expresses a differential effect.
Specification of the bulge and disk spectral energy distributions is needed
when one investigates how attenuation by internal dust
affects the integrated photometry of a bulge$+$disk system as a whole
(e.g. Pierini et al. 2003; see also Tuffs et al. 2004).
In fact, in this case one needs to count the photons
produced either in the bulge or in the disk.

Finally we note that the different curves reproduced
in the following figures are not completely smooth and monotonic,
from one wavelength included in the models to the next one,
because of photon noise in the Monte Carlo simulations.

\subsection{The face-on case: effects due to the disk opacity}

\subsubsection{The attenuation function}

\subsubsubsection{Bulge}

Fig. 1 reproduces the attenuation function calculated for the bulge
of a late-type galaxy (like our own) seen face on, as a function of $\tau_V$
(half the total central dust column density of the disk)
and the structure of the dusty ISM.
It also shows the extinction function expected for a homogeneous,
non-scattering MW-type dust screen foreground to the light source
with total V-band extinction optical depth equal to the total
attenuation optical depth at V band of the face-on bulge model
with a homogeneous dust disk reproduced in each panel.
This normalization is chosen in order to better show the difference in shape
between the curves reproduced in Fig. 1.
The comparison between these different curves
reminds the non-specialized reader, if needed,
that the dust obscuration for an extended object like a bulge
is conceptually different from the dust extinction of stars
(see footnote 1 and e.g. Calzetti 2001).

Fig. 2 reproduces the same set of attenuation functions
displayed in Fig. 1, each one normalized to the value of the attenuation
at 0.55 $\rm \mu m$ (i.e., $A_V$).
In origin, this normalization allowed the comparison of the shapes
of extinction curves with different columns of dust at any wavelength.
Here it is used to emphasize two aspects of the attenuation function:
its slope in the far-UV spectral region
and the presence of the so-called ``extinction bump'' at $\rm \sim 2175~\AA$,
which is a signature of the average MW-type dust
(see Whittet et al. 2004 for a recent discussion) that is used in our models.

In addition, in Fig. 2 we reproduce the extinction function normalized
to its V-band value expected for a homogeneous, non-scattering MW-type
dust screen foreground to the light source, whatever the total V-band
extinction optical depth.
This normalized extinction function is equal to the MW extinction curve
normalized to its V-band value listed in Table 1.
The difference in shape between the curves reproduced in Fig. 2 is evident.

From the visual inspection of Fig. 1 and 2,
the following characteristics emerge.
\begin{itemize}
\item
The average slope of the attenuation function for a face-on bulge
increases together with the opacity, for $\tau_V \le 1$,
whatever the structure of the dusty ISM.
Hence, for the face-on bulge, dust reddening increases
together with the disk opacity within the wavelength range
where at least the inner region of the dust disk is optically thin.
\item
The average slope of the attenuation function flattens (steepens)
at wavelengths shorter (larger) than $\rm \sim 0.55~\mu m$
when the opacity increases, for $\tau_V > 1$,
whatever the structure of the dusty ISM.
Hence, for the face-on bulge, dust reddening tends to a constant value
within the wavelength range where at least the inner region
of the dust disk becomes optically thick, owing to an increase in opacity.
\item
For a fixed $\lambda$, $A_{\lambda}$ increases with increasing opacity,
the rate of increase diminishing rapidly at wavelengths shorter than
$\rm \sim 0.3~\mu m$ for $\tau_V \le 2$, whatever the structure of
the dusty ISM.
As a consequence, from the optical through the far-UV spectral range,
$A_{\lambda}$ is almost the same for $2 \le \tau_V \le 8$, so that
the maximum attenuation (at $\lambda = \rm 0.1~\mu m$) increases
only from 0.5 to 1.4 mag when $\tau_V$ increases from 0.25 to 8.
Hence, for the face-on bulge, the increase in dust attenuation
as a function of the disk opacity is strongly non-linear.
\item
For a fixed value of $\tau_V $ (especially for $\tau_V \ge 1$),
the structure of the dusty ISM does not affect to zeroth order
the shape of the attenuation function for the face-on bulge.
\item
Nevertheless, a homogeneous dust distribution produces a larger attenuation
than a two-phase clumpy one, whatever $\tau_V $ (for $\tau_V \le 8$).
This difference reaches a maximum factor of $\sim$ 1.6 for $\tau_V \sim 2$.
\item
For very low values of the opacity (i.e., for $\tau_V \sim 0.25$),
the dust distribution, be it homogeneous or two-phase clumpy,
has an almost negligible impact on the attenuation function
for the face-on bulge.
Slight discrepancies (of the order of 0.15 magnitudes or less)
are present only shortward of $\lambda \rm \sim 0.3~\mu m$
(i.e., rightward of $1/\lambda \rm \sim 3.3~\mu m^{-1}$).
\item
At wavelengths shorter than $\rm \sim 0.3~\mu m$,
the discrepancy between the homogeneous case and the two-phase clumpy case
increases with increasing opacity, for $\tau_V \le 2$,
but decreases with increasing opacity, for $2 < \tau_V \le 8$.
\item
Conversely, at wavelengths longer than $\rm \sim 0.55~\mu m$
(i.e., leftward of $1/\lambda \rm \sim 1.8~\mu m^{-1}$),
the discrepancy between the homogeneous case and the two-phase clumpy case
increases with increasing opacity along the whole range of $\tau_V$
explored by the models.
\item
The relative prominence of the $\rm 2175~\AA$-extinction bump diminishes
when $\tau_V$ increases.
Thus, radiative transfer effect may weaken this feature
in far-UV spectra of face-on bulges of nearby galaxies.
\end{itemize}
This behavior may be explained as follows.

First we recall that, by construction (see Sect. 2), the face-on projection
of the bulge is almost coincident with the face-on projection
of the dustiest region of the disk (within one scale length),
where the {\it local} optical depths are thus the largest.
For this projection, almost half of the bulge stands between the dust disk
and the outside observer.

The light produced in the face-on bulge is effectively removed
from the observer's line of sight when the amount of dust increases,
especially when it is emitted at $\lambda \rm \le 0.3~\mu m$.
However, the increase of $A_{\lambda}$ together with $\tau_V$
slows down when a {\it local} optical depth of order 1 is approached
across the disk region corresponding to the face-on projection of the bulge
(first at the shortest wavelengths and then at progressively larger ones).
In the ideal case of a bulge bisected by an opaque dust layer,
we expect that this dust blocks {\it exactly} half of the light
emitted by the bulge towards the outside observer
(i.e., $A_{\lambda} = 0.75$ for any $\lambda$),
when we look at it from a direction perpendicular to the dust layer.
Obviously, this is not the dust distribution of our models,
so that the shape of the attenuation function for the face-on bulge
changes as a function of $\tau_V$, while $A_{\lambda}$ approaches
a sort of asymptotic value, when $\tau_V$ increases from 0.25 to 8.
This causes the simultaneous flattening of the attenuation function,
from the shortest wavelengths towards progressively longer ones,
for $\tau_V > 1$, whatever the structure of the dusty ISM of the disk.
The asymptotic value is larger than 0.75 since the dust/stars configuration
of our model does not resemble the ideal case previously described.

The simulations for the face-on bulge confirm that
both absorption and scattering diminish in a two-phase clumpy medium,
in comparison with a homogeneous medium having the same total mass and type
of dust (see Witt \& Gordon 1996 for a detailed analysis).
It is for this reason that the attenuation of the light
produced in the face-on bulge at a given wavelength becomes ``grey''
(i.e., tends to an asymptotic value) at slightly lower values of $\tau_V$
for a homogeneous dust distribution than for a two-phase clumpy one.
{\it However, the structure of the dusty ISM of the disk
produces only negligible effects on the observed colors of a face-on bulge.}

As a further result of the radiative transfer calculations,
we find that the absorption excess represented
by the $\rm 2175~\AA$-extinction bump (Calzetti et al. 1995)
is washed out when $\tau_V$ increases.
This effect was noted first by Cimatti et al. (1997)
for a different dust/stars configuration from that assumed here.
What happens in Fig. 1 and 2 is that, for photons of the bulge
emitted at neighbouring wavelengths with respect to $\rm 2175~\AA$,
chances for both (multiple) scattering and the eventual absorption
increase together with $\tau_V$,
so that the attenuation reaches an asymptotic value
at and around $\rm 2175~\AA$.

\subsubsubsection{Disk}

Fig. 3 and 4 reproduce the attenuation function
and the normalized attenuation function, respectively,
for the disk of a late-type galaxy seen face on, as a function of
the opacity and the structure of the dusty ISM of the disk,
in analogy with Fig. 1 and 2, respectively, for the bulge.

From the visual inspection of Fig. 3 and 4, we draw the following conclusions.
\begin{itemize}
\item
The average slope of the attenuation function for a face-on disk
increases with increasing opacity, through the whole range of $\tau_V$
considered here, whatever the structure of the dusty ISM.
Hence, for the face-on disk, dust reddening is a monotonically increasing
function of the opacity.
\item
However, the attenuation optical depth increases at a reduced rate,
at least in the UV spectral region, when $\tau_V$ becomes larger than 1,
so that the maximum attenuation (at $\lambda = \rm 0.1~\mu m$)
increases from 0.2 to 2.6 mag (for the homogeneous dust distribution)
when $\tau_V$ increases from 0.25 to 8.
\item
A difference in the shape of the attenuation function persists
between the homogeneous case and the two-phase clumpy case,
whatever $\tau_V$ (for $\tau_V \le 8$).
\item
Also for a face-on disk the homogeneous dust distribution produces
a larger attenuation than a two-phase clumpy one, at any wavelength.
This time the discrepancy in the values of $A_{\lambda}$ produced
by the two very different structures of the dusty ISM increases
with increasing total central dust column density, whatever the wavelength.
\item
The absorption-excess feature at $\rm \sim 2175~\AA$
can be easily identified, though its prominence diminishes
with increasing opacity (for $\tau_V \le 8$).
\end{itemize}

In order to understand this behavior, it is necessary to refresh first
some properties of the dust/stars configuration of our disk models.
The distribution of dust and stars along the radial coordinate of the disk
is assumed to be described by the same exponential law,
whatever the parent stellar population of the stars.
Conversely, the vertical distribution of the different stars of the disk
depends on their parent stellar population.
Thus the scale height decreases together with the wavelength
characterizing to zeroth order the age of the individual stars
(see Sect. 2 and Table 1).
As a consequence, the fraction of light which is produced in the disk
at a given wavelength and is affected by the dust,
depends even more on $\lambda$, with respect to the case
of a vertical distribution with a unique scale height
(as e.g. in Ferrara et al. 1999), for fixed extinction properties,
total amount of dust, and structure of the dusty ISM of the disk.

Having assumed the dust/stars configuration for the disk as in Sect. 2,
we note that the light produced by the stars in the outskirts of the disk
will shine almost unattenuated for very low values of the disk opacity,
whatever the wavelength.
However, when $\tau_V$ increases, a {\it local} optical depth $\sim 1$
or larger will be eventually reached within a progressively thicker region,
delimited by a progressively larger radius;
the longer $\lambda$ the smaller this region.
It is now straightforward to understand why, for the face-on disk,
$A_{\lambda}$ increases together with $\tau_V$, at any given $\lambda$,
and why this increase is so dramatic at $\lambda \le \rm 0.3~\mu m$
(see Table 1).

The best example of the importance of the dust/stars configuration
for the result of the radiative transfer analyzed here
is offered by the far-UV photons emitted at $\lambda \le \rm 0.18~\mu m$,
since they are produced by the stars the most deeply embedded in the dust layer
(see Table 1).
These photons might eventually never reach the outside observer
if $\tau_V$ were allowed to increase indefinitely,
whatever the height from the galactic plane and radius where they are produced.
Conversely, optical/near-IR photons may still reach the observer
in large number even for $\tau_V = 8$.

For photons emitted at a fixed wavelength, a constant value of the attenuation
equal to 0.75 mag is expected when we see half the light from the face-on disk
bisected by an opaque dust layer.
Since the dust/stars configuration matters, our face-on disk models
produce an attenuation of less than 0.75 mag for photons emitted shortward of
$\lambda = \rm 0.55~\mu m$ even for $\tau_V = 8.0$.
We also note that the attenuation at UV wavelengths almost doubles
its initial value when $\tau_V$ increase from 0.25 to 0.5, and from 0.5 to 1.
At those wavelengths, the increase of $A_{\lambda}$
together with $\tau_V$ slows down for larger values of $\tau_V$.
Despite that, the absorption-excess feature at $\rm \sim 2175~\AA$
may still be easily identified in the attenuation function
for the face-on disk model with $\tau_V = 8$,
though its prominence has become modest.
Information on this feature is carried by the light escaping from the regions
of the disk where the {\it local} optical depth is low enough
(e.g. in the outskirts of the disk).

Finally, we note that the attenuation in the K band assumes
a slightly negative value in the case of a face-on disk
with $\tau_V \sim 0.25$ (or less).
This means that the outside observer receives more K-band photons
from the face-on disk than it would be in absence of dust.
The reason for this excess is the contribution of those K-band photons
initially traveling in different directions but scattered by the dust
towards the outside observer.
This effect has already been highlighted by Baes \& Dejonghe (2001).

From the visual inspection of Fig. 1 through 4 it is clear that
the solutions of the radiative transfer for bulge and disk
of a late-type galaxy seen face on are different.
The reason is the different distribution of stars and dust for either system
with respect to the outside observer.
The face-on projection of the bulge is always within the dustiest region
of the disk (by construction), and, if the opacity is large enough,
the fraction of stars whose light is significantly affected by the dust
is the same whatever the emission wavelength.
Conversely, for the same value of the opacity,
the face-on projection of the disk will present regions almost unaffected
by the dust at larger galactocentric distances
the shorter the wavelength of the emitted photons.
At the same time, the fraction of stars of the disk
whose light is significantly affected by the dust is larger
the shorter the emission wavelength, whatever the radial distance
(also by construction).

Of course, for both bulge and disk, geometry is convolved
with the selective nature of absorption and scattering
in the computation of the attenuation function.

\subsubsection{The color excess}

Here we discuss the behavior of the optical color excess $E(B-V)$
and the near-IR color excess $E(J-K)$, as defined in Sect. 2,
as a function of the opacity and the structure of the dusty ISM of the disk.
This exercise offers a very useful example of how
confusing the extinction law of the mixture of dust grains
present in the dusty ISM of a galaxy with the attenuation function
of the galaxy itself may lead to extremely wrong conclusions
regarding the color excess expected for this dusty stellar system.
Furthermore, we discuss whether the ratio of total to selective
{\it attenuation} $R_V=A_V/E(B-V)$ for a dusty stellar system,
like a late-type galaxy, represents a diagnostic of the total-to-selective
{\it extinction} ratio of the mixture of dust grains that this system contains.

The values of the attenuation function at the effective wavelengths
of the filters B, V, J, and K (i.e., at $\lambda =$ 0.44, 0.55, 1.25,
and $\rm 2.2~\mu m$, respectively) are determined
from the linear interpolation of the models displayed in Fig. 1 and 3.

\subsubsubsection{Bulge}

For the face-on bulge, $E(B-V)$ reaches an absolute maximum
of about 0.1 mag for $\tau_V \sim 2$ and decreases down to $\sim 0.05$ mag
for larger values of the opacity in the homogeneous case,
while it never exceeds $\sim 0.06$ mag (for $\tau_V \le 8$)
in the two-phase clumpy case (see Fig. 5a).
Conversely, $E(J-K)$ increases non-linearly with $\tau_V$ up to $\sim 0.29$ mag
or $\sim 0.2$ mag, for the homogeneous or two-phase clumpy dust distribution,
respectively, the increase being slower in the latter case (see Fig. 5b).
Note that, for both local distributions of the dust,
$E(J-K)$ is definitely lower than $E(B-V)$ only for $\tau_V \le 1$.

This behavior is surprising only at first glance.
In fact we can understand it when considering the results
reproduced in Fig. 1 and 2.
There it is shown that $A_B$ is slightly greater than $A_V$
and increases faster than $A_V$ for $\tau_V < 2$, especially
in the homogeneous case.
However, $A_B - A_V \rm \rightarrow 0$ for larger values of the opacity,
when the number of lines of sight with optical depths at B- and V-band
less than unity is very much reduced across the region of the dust disk
crossed by photons originated in the bulge.
This regime is reached at a slightly lower value of $\tau_V$
(where the turn-around of $E(B-V)$ takes place)
for the homogeneous dust distribution than for the two-phase clumpy one.
$E(B-V)$ never reaches values larger than 0.1 mag when $\tau_V$ increases
beyond $\sim 2$ since the attenuation function at optical wavelengths
is rather flat for $\tau_V \ge 2$.
Thus, we conclude that the amount of reddening of a (face-on) bulge,
as measured by $E(B-V)$, is not a measure of the opacity
of the associated dust disk, except for low values of $\tau_V$.

Conversely, Fig. 1 and 2 show that $A_J$ is greater than $A_K$,
and increases faster than $A_K$ for $\tau_V \le 4$.
However, the attenuation function tends to flatten even in the near-IR
for $\tau_V > 4$.
As a consequence, $E(J-K)$ is a monotonically increasing function of $\tau_V$,
for the whole range of opacities sampled here, but it is expected
to show a behavior similar to that of $E(B-V)$ if even larger values
of the opacity were considered.

Furthermore, Fig. 1 and 2 show that the attenuation function
of a face-on bulge is steeper in the near-IR than in the optical,
for $\tau_V \ge 2$.
For this reason, $E(J-K)$ may reach larger values than $E(B-V)$.
The decrease of $E(B-V)/E(J-K)$ towards larger values of $\tau_V$ (see Fig. 6a)
is the consequence of the previously described behavior
of the attenuation optical depth at each of the four effective wavelengths
as a function of the amount of dust.

Fig. 6b shows another result which is surprising only at first glance:
for the face-on bulge, $R_V$ spans a wide range of values
(from $\sim 3$ to $\sim 19$ and from $\sim 3$ to $\sim 13$,
in the homogeneous and two-phase clumpy cases, respectively)
as a function of $\tau_V$.
As a result of the radiative transfer, $R_V$, here defined as
the ratio of the {\it attenuation} at $\rm 0.55~\mu m$
to the color excess $E(B-V)$, significantly departs from the value of 3.1,
which is the total-to-selective {\it extinction} ratio
for the average MW-type extinction law (see e.g. Valencic et al. 2004).
The non-linear dependence on the total amount of dust of $A_V/E(B-V)$ reflects
the flattening of the attenuation function with increasing $\tau_V$ (Fig. 2).

\subsubsubsection{Disk}

For the face-on disk, $E(B-V)$ and $E(J-K)$ are monotonically increasing
functions of the opacity, at least for $\tau_V \le 8$ (see Fig. 7).
However, we note that a variation of $E(B-V)$ as small as 0.02 mag
may correspond to a variation of $\tau_V$ by a factor of 2.
Thus we conclude that also the amount of reddening of a face-on disk,
as measured by $E(B-V)$, is not a sensitive measure of the opacity of the disk.

Fig. 7 also shows that the optical and near-IR color excesses
are larger in the homogeneous case than in the two-phase clumpy one.
Furthermore, $E(B-V)$ is greater (lower) than $E(J-K)$
for $\tau_V < \sim 5$ ($\tau_V > \sim 5$).
As a consequence, $E(B-V)/E(J-K)$ drops from $\sim 10$ to the asymptotic
value of $\sim 0.8$ (see Fig. 8a), the dynamic range of this drop
being similar to that observed for the face-on bulge
(from $\sim 2.5$ to $\sim 0.25$) in Fig. 6a.
Once again, the behavior of $E(B-V)/E(J-K)$ for the face-on disk
reflects the change of the shape of the attenuation curve
as a function of $\tau_V$ (Fig. 4).

Finally, for the face-on disk, $R_V$ increases from $\sim 1.8$
up to the asymptotic value of $\sim 5$ ($\sim 6$)
in the homogeneous (two-phase clumpy) case,
when $\tau_V$ increases from 0.25 to 8 (Fig. 8b).
This is a consequence of the steepening of the attenuation function
in the optical spectral region with increasing opacity.

From Fig. 6b and 8b we conclude that radiative transfer effects
hinder the evaluation of the total-to-selective {\it extinction} ratio,
typical of the dust mixture present in a dusty system
like a bulge$+$disk galaxy, from the total-to-selective {\it attenuation} ratio
observed for either bulge or disk.

\subsubsection{The importance of the scattered light}

Here we discuss in detail the fraction of light, emitted at a given wavelength
(from the far-UV through the near-IR) in the bulge or disk component
of a face-on late-type galaxy, which is either scattered by dust
or directly transmitted towards the outside observer.

\subsubsubsection{Bulge}

For the face-on bulge, the fraction of light scattered
towards the outside observer ($f(sca)$) never exceeds 20 per cent of
the light initially emitted along the observer's line of sight,
for the range of opacity investigated here (see Fig. 9).
The wavelength where this fraction peaks increases
when $\tau_V$ increases.
Shortward of this peak wavelength, the fraction of scattered light decreases
with increasing disk opacity.
As a consequence, the color of the scattered light is blue for
$\tau_V \le 0.5$ and red for larger values of $\tau_V$.
This behavior may be easily understood when considering that
the efficiencies of both absorption and (multiple) scattering
at a given wavelength increase when $\tau_V$ increases.
When $\tau_V$ increases, (multiple) scattered far-UV photons
originated in the face-on bulge will be eventually absorbed
by the face-on dust disk; conversely, optical/near-IR photons originated
in the face-on bulge will be scattered more efficiently
in the direction to the outside observer, initially,
but absorbed in increasing proportion as soon as the {\it local} optical depths
at these wavelengths equal unity or become larger
across the region of the dust disk crossed by those photons.

Conversely, the fraction of light directly transmitted
towards the outside observer ($f(dir)$) is a monotonically decreasing function
of $\tau_V$, whatever the wavelength (see Fig. 10).
This fraction becomes constant shortward of a break wavelength
for $\tau_V > 1$, the break wavelength increasing
with increasing values of the opacity.
Longward of this wavelength the fraction of light directly transmitted
along the observer's line of sight is larger by up to a factor of two.

As a consequence of the previous behavior, the scattered-to-directly
transmitted light ratio ($f(sca)/f(dir)$) increases (decreases)
with decreasing wavelength longward (shortward) of the break wavelength
(see Fig. 11).
This ratio never exceeds 35 per cent at any given $\lambda$.

Finally, we find that the clumpiness of the dusty ISM
reduces both absorption and scattering of the light
emitted by the face-on bulge, whatever the amount of dust.
However, for a face-on bulge, $f(sca)/f(dir)$ is only mildly
(if not negligibly) dependent on the clumpiness of the dusty ISM of the disk.

\subsubsubsection{Disk}

For the face-on disk, $f(sca)$ increases together with $\tau_V$,
for $\tau_V \le 2$, while it decreases when $\tau_V$ becomes larger,
whatever the wavelength (see Fig. 12).
The peak of the scattered light is in the far-UV spectral region
for values of $\tau_V$ as large as 4; it moves to the optical/near-IR
spectral region for larger values of $\tau_V$.
The interpretation of this behavior is the same as for the face-on bulge.

Conversely, $f(dir)$ is a monotonically decreasing function of $\tau_V$,
whatever the wavelength, and no break wavelength is present (see Fig. 13).
This may be easily understood from the discussion in Sect. 3.1.1.

In addition, $f(sca)/f(dir)$ increases together with the opacity,
whatever the wavelength.
This is due to the dramatic decrease of $f(dir)$ when $\tau_V$ increases,
especially for the far-UV wavelengths.
The increase of $f(sca)/f(dir)$ as a function of $\tau_V$
strongly depends on the clumpiness of the dusty ISM (see Fig. 14).
Furthermore, $f(sca)/f(dir)$ peaks at about 0.15 $\rm \mu m$
whatever $\tau_V$ (for $\tau_V \le 8$).
We note that, for $\tau_V = 8$, $f(sca)$ is about 50 (30) per cent of
the 0.15 $\rm \mu m$-flux reaching the observer in the homogeneous
(two-phase clumpy) case.

There is a lot of difference concerning the behavior of $f(sca)/f(dir)$
between the face-on disk and the face-on bulge.
Once again the reason is the different configuration of dust and stars
emitting at a given wavelength for the bulge and the disk.

\subsection{The $\bf \tau_V =1$ case: effects due to the inclination}

\subsubsection{The attenuation function}

\subsubsubsection{Bulge}

Fig. 15 shows the complex behavior of $A_{\lambda}$ for the bulge of a galaxy
with a dust disk having $\tau_V = 1$ and MW-type dust,
as a function of the inclination of the galaxy.
The attenuation optical depth increases faster at optical wavelengths
than at far-UV ones, when $i$ increases from $\rm 0^o$ to $\rm 70^o$,
as can be realized from Fig. 16, where $A_{\lambda}/A_V$ is displayed.
The final result is that the attenuation function flattens
shortward of $\rm \sim 0.66~\mu m$ when $i = \rm 70^o$.
At the same wavelengths, the increase is reversed
and the flattening accentuated for $i$ $\rm > 70^o$.
Conversely, longward of $\rm \sim 0.66~\mu m$, the attenuation optical depth
increases proportionally to $1/\lambda$, with the result that
the attenuation function steepens, when $i$ increases
from $\rm 0^o$ to $\rm 70^o$.
For higher inclinations, the attenuation optical depth increases faster
at the largest near-IR wavelengths than at the shortest ones,
so that $A_{\lambda}$ tends to flatten also in the near-IR.

When the inclination of the galaxy becomes larger,
a photon of given wavelength, emitted in the bulge
and traveling through the dust disk on its way to the outside observer,
faces a larger number of lines of sight with optical depths greater than unity
whatever $\lambda$.
On the other hand, a larger fraction of the bulge comes into the view
of the outside observer unobstructed by the dust disk, when $i$ increases.
In fact, the bulge model has a (wavelength-independent) scale height
larger than that of the dust distribution by almost a factor of 10,
and a radial extension 3 times shorter than that of the dust disk
(cf. Sect. 2).
Fig. 15 shows that the increase of the projected surface of the bulge
which is unobstructed by the dust disk prevails over the competing increase
of the {\it local} optical depth through the dust disk, for all wavelengths
except the longest near-IR ones.
At the same time, the attenuation function for the edge-on bulge is grey
throughout almost the whole spectral region under consideration.
For this reason, the absorption peak at $\rm \sim 2175~\AA$,
typical of the MW-type extinction law, is almost undetectable
in the attenuation function of the edge-on bulge.

From the previous discussion, it is clear that the blocking action
of the dust disk must peak at some intermediate inclination,
since almost half of the bulge stands between the dust disk and the observer,
in the face-on case, and most of the projected surface of the bulge
is unobstructed by the dust disk, in the edge-on case.
The value of the inclination which maximizes the blocking action
of the dust disk will depend on the wavelength of the emitted photons.
In addition to the previous two geometric effects,
the wavelength-dependence of the scattering phase function
asymmetry parameter (see Table 1) plays a role in interpreting Fig. 15 and 16.
We recall that scattering is more forward directed the shorter the wavelength
of the scattered photon.
Thus the near-IR photons emitted in the bulge
and traveling through the dust disk will have a greater probability
of being scattered out of the observer's line of sight
than the optical/far-UV ones, in the edge-on case.
This contributes to explaining why the attenuation by internal dust
of the light produced in the bulge at near-IR wavelengths
increases monotonically together with the inclination of the galaxy.

Finally, we note that the clumpiness of the dusty ISM of the disk
(with $\tau_V =1$) produces distinguishable effects on the attenuation function
for the bulge only for the lowest inclinations of the galaxy,
i.e., when $A_{\lambda}$ is the lowest whatever $\lambda$.
This is explained in the following section.

\subsubsubsection{Disk}

The dependence on inclination of the attenuation function
for a disk with $\tau_V = 1$ and MW-type dust is very dramatic,
as shown in Fig. 17.
In fact, the attenuation increases (almost doubling its initial value)
at any $\lambda$, when $i$ increases from $\rm 0^o$ up to $\rm 70^o$,
owing to the increasing probability of absorption
for a photon of any wavelength traveling through the dust disk
towards the outside observer.
For higher inclinations, the attenuation function becomes suddenly very steep,
with the result that $A_{\lambda} \sim 0.4 \times 1 / \lambda$
in the edge-on case.

For inclinations close to $\rm 90^o$, the photons produced in the dustiest
region of the disk and traveling towards the outside observer
will be eventually absorbed there, whatever their wavelength.
The photons produced in the peripheral (and less dusty) regions
of the disk and traveling towards the outside observer
will experience significantly increased values
of the line-of-sight optical depth with respect to the face-on case.
They may be scattered out of the observer's line of sight,
if their wavelength belongs to the optical/near-IR domain.
Conversely, photons produced at higher heights above the disk
will have a greater probability to reach the outside observer.
Since the ratio of stellar disk scale-height and dust disk scale-height
is an increasing function of the emission wavelength, by construction
(see Sect. 2 and Table 1), absorption will be even more dramatic
the shorter the wavelength.
In addition to the previous geometrical effect,
the wavelength-dependence of the scattering phase function asymmetry parameter
(see Table 1) predicts that (multiple) scattered photons
will be eventually absorbed in larger fraction the shorter their wavelength,
as already discussed.
This explains the abrupt rise of $A_{\lambda}$ at far-UV wavelengths
for inclinations close to $\rm 90^o$.
As a consequence, for these inclinations, the $\rm 2175~\AA$-bump,
typical of MW-type dust, is almost undetectable (see Fig. 17 and 18).

Finally, we note that the impact of the structure of the dusty ISM
on the attenuation function, in terms of absolute values and shape,
is almost negligible when the disk is seen at $i > \rm 70^o$.
For such high inclinations, the optical depth
along the observer's line of sight may be much larger than unity
for many if not most of the wavelengths just because many dust clumps
line up in that direction.
Hence the eventual absorption of a photon of given wavelength
happens just by virtue of the total cross section (or blocking effect)
of the dust clumps.
This result complements that of Matthews \& Woods (2001),
who noted that differences in the structure of the dusty ISM,
for their edge-on models, impact sensitively on the values of the attenuation
but not on the shape of the attenuation function.

\subsubsection{The color excess}

\subsubsubsection{Bulge}

For the bulge models associated with a dust disk having $\tau_V = 1$,
Fig. 19a shows that either $E(B-V)$ is constant (homogeneous dust distribution)
or it slowly increases together with the inclination (two-phase clumpy medium),
for $\rm 0^o \le $ $i$ $\le \rm 70^o$.
However, $E(B-V)$ declines for higher inclinations,
whatever the local distribution of the dust.
As a consequence, for the bulge $E(B-V)$ is bluer (by 0.04 mag)
in the edge-on case than in the face-on case.

Conversely, $E(J-K)$ is a strongly non-linear function of
the galaxy inclination: it increases from $\sim 0.04$ to $\sim 0.2$ mag,
when $i$ increases from $\rm 0^o$ to $\rm \sim 80^o$;
it decreases for higher inclinations, whatever the dust distribution
(see Fig. 19b).
The initial increase of $E(J-K)$ is a consequence of the steepening of
the attenuation function for the bulge in the near-IR,
when $i$ increases from $\rm 0^o$ to $\rm 70^o$,
which means that the dusty medium crossed by the photons emitted in the bulge
and traveling towards the outside observer tends to become optically thick
also in this wavelength domain.

The decrease of both $E(B-V)$ and $E(J-K)$
for inclinations approaching $\rm 90^o$ is due to the fact that
an increasing fraction of the light produced in the bulge
reaches the outside observer without being affected by the dust
(within the disk).
The turn-over in the behavior of the color excess
happens when the increase in the line-of-sight optical depth
faced by the photons traveling through the dusty medium
is compensated by the reduction in the number of such photons.
It occurs at higher inclinations for $E(J-K)$ than for $E(B-V)$
because of the wavelength-dependence of the extinction law (see Table 1).

As a consequence of the previous behavior,
$E(B-V)/E(J-K)$ decreases monotonically from 1.7 to 0.3
when $i$ increases from $\rm 0^o$ to $\rm 90^o$ (see Fig. 20a).
Thus, for the bulge, $E(J-K)$ is almost three times larger than $E(B-V)$
for $i$ $\rm \sim 80^o$.

Conversely, $R_V$ is almost constant ($\sim 4$), for $i \le \rm 40^o$,
but increases for higher inclinations (up to $\sim 12$ for the edge-on bulge)
without any significant dependence on the local distribution of the dust
(see Fig. 20b).

\subsubsubsection{Disk}

For a disk with $\tau_V = 1$, both $E(B-V)$ and $E(J-K)$
are monotonically increasing functions of $i$,
$E(B-V)$ being larger than $E(J-K)$ for $i$ $\le \rm 80^o$ (see Fig. 21).
This is a consequence of the fact that
$A_\lambda$ increases monotonically together with $i$ (cf. Fig. 17),
the increase being slower in the optical than in the near-IR
when inclinations higher than $\rm 70^o$ are approached.
In addition, we find that the clumpiness of the dusty ISM
does not affect the behavior of the near-IR color excess,
but produces a slight difference for the optical one, as expected from Fig. 17.

As a consequence of the previous behavior,
$E(B-V)/E(J-K)$ is a monotonically decreasing function of $i$,
ranging from $\sim 3$ (face-on disk) to $\sim 1$ (edge-on disk),
as shown in Fig. 22a.
This range is similar to that obtained for the bulge (cf. Fig. 20a).

Conversely, $R_V$ is a slowly increasing function of $i$ (see Fig. 22b),
ranging from $\sim 3$ (face-on disk) to $\sim 6$ (edge-on disk).
This confirms that, for the disk, the attenuation at optical wavelengths
increases slightly faster than the reddening at the same wavelengths,
even for $i \ge \rm 70^o$ (cf. Fig. 17 and 18).
This is due to the geometrical increase of the region of the disk
which is opaque to optical photons.

\subsubsection{The importance of the scattered light}

\subsubsubsection{Bulge}

The fraction of light which is produced in the bulge and is scattered
by the dust disk towards the outside observer diminishes
when the inclination of the galaxy increases (see Fig. 23).
It ranges from $\sim$ 20 per cent (face-on case) to 5 per cent (edge-on case).
For the case of $\tau_V = 1$ considered here, the color of the scattered light
is always red and the clumpiness of the dusty ISM does not play
an important role.

Conversely, the fraction of light which is produced in the bulge
and is unaffacted by the dust (and, thus, reaches directly
the outside observer) has a non-monotonic behavior
as a function of the inclination of the galaxy (see Fig. 24).
For $i \le \rm 70^o$, this fraction decreases with increasing $i$,
whatever $\lambda$.
For higher inclinations, a minor fraction of the dust disk projects
into the bulge, so that $f(dir)$ increases again in the optical/far-UV
spectral domain, where absorption already prevails over scattering
for $\tau_V =1$.
The reason why $f(dir)$ keeps on decreasing in the near-IR,
for $i > \rm 70^o$, although a larger fraction of the bulge
comes into the view of the outside observer without being obstructed
by the dust disk, is that both absorption and scattering
out of the observer's line of sight become more effective in the near-IR,
for $\tau_V = 1$.

The effect on $f(dir)$ of the clumpiness of the dusty ISM
diminishes for increasing inclinations,
owing to the blocking effect of the dust clumps
along the observer's line of sight, so that the two model curves in Fig. 24
are undistinguishable for $i$ $> \rm 70^o$.

As a result of the behaviors previously described,
for the bulge $f(sca)/f(dir)$ dimishes from 35 per cent (at most)
to 15 per cent (at most) when the inclination of the galaxy increases
from $0^o$ to $\rm 90^o$ (see Fig. 25).

\subsubsubsection{Disk}

When the inclination of the galaxy increases from $0^o$ to $\rm 90^o$,
the photons produced in the disk and traveling through its dusty ISM
will experience larger line-of-sight optical depths
on their way towards the outside observer,
simply because the average path length within the dust distribution increases.
This implies that the probability of multiple scattering events rises
for increasing inclinations, as well as the probability that
the multiple-scattered photons are eventually absorbed inside the dust disk.
Thus, $f(sca)$ and $f(dir)$ are expected to monotonically decrease
for any $\lambda$ when $i$ increases, as shown in Fig. 26 and 27, respectively.

We note that the color of the scattered light turns from blue to red
only for the highest inclinations, when the far-UV photons are absorbed
for the most part.
Nevertheless, in the edge-on case, the scattered photons
make up to 30 per cent of the total far-UV photons
which reach the outside observer (see Fig. 28).
In the optical and near-IR, $f(sca)/f(dir)$ slightly increases for
$\rm 0^o \le$ $i$ $\rm \le 70^o$, but drops for higher inclinations,
when absorption prevails over scattering along the observer's line of sight,
and the scattered photons which manage to leave the system
do so mostly in other directions or from the peripheral regions of the disk,
where the {\it local} optical depth is low.

\section{Summary and Discussion}

The interpretation of the spectral energy distribution
observed for a dusty stellar system requires understanding
the radiative transfer problem for that particular system.
On the other hand, the solution given by a radiative transfer model
(e.g. the attenuation function, investigated here, which describes
the behavior of the total attenuation optical depth
as a function of wavelength) depends on the dust/stars configuration,
the total amount of dust, the extinction properties of the mixture
of dust grains (e.g. Gordon et al. 2003), and the structure
of the dusty interstellar medium (ISM) assumed for the system.

These aspects have been extensively discussed
by e.g. Witt \& Gordon (1996, 2000) for the spherically symmetric case.
Different descriptions of the dust/stars configuration
and/or the structure of the dusty ISM are adopted
by the existing different calculations of the attenuation function
(including multiple scattering) for a bulge$+$disk galaxy,
whether these computations are performed analytically (e.g. Byun et al. 1994;
Silva et al. 1998; Xilouris et al. 1999; Baes \& Dejonghe 2001;
Tuffs et al. 2004) or via Monte Carlo techniques (e.g. de Jong 1996; Wood 1997;
Kuchinski et al. 1998; Ferrara et al. 1999; Bianchi et al. 2000;
Matthews \& Wood 2001).
Some of these authors discuss in detail the further dependences
of the attenuation function on the inclination,
the structure of the dusty ISM (Bianchi et al. 2000; Matthews \& Wood 2001),
and the bulge-to-disk ratio of the galaxy (Ferrara et al. 1999;
Tuffs et al. 2004).

We present new Monte Carlo simulations for the attenuation by internal dust
of the light produced either in the bulge or in the disk
of a giant late-type galaxy like the Milky Way, based on the DIRTY code
of radiative transfer of stellar and scattered radiation (Gordon et al. 2001).
These models do not include the absorption and scattering of photons
associated with all recombination lines
{\it in H {\small II} regions themselves},
as do those of Silva et al. (1998), Panuzzo et al. (2003),
and Tuffs et al. (2004).
However they do include the absorption and scattering of photons
associated with all recombination lines {\it after they emerge
from the H {\small II} regions}.
In addition, they explore a different and important region
of the parameter space, concerning both the description of the clumpiness
of the dusty ISM of the disk, and the assumed 3-D spatial distribution
of the sources of photons of fixed wavelength $\lambda$
with respect to the 3-D spatial distribution of the dust.

These aspects are discussed in Sect. 2.
Here we recall that our models assume a purely stochastic distribution
of the dust clumps, with respect to each other and the bluest stars.
The dust clumps are simulated as cubic cells with a size of 44 pc,
and a filling factor set equal to 0.15 all over the dust disk,
while the clumpy-to-diffuse ISM phase density ratio is set equal to 100.
This set of parameters was found by Witt \& Gordon (1996) to be closest
to that representing the structured ISM of the Galaxy.
For a spherically symmetric geometry, it represents an intermediate case
between the case where the interclump medium controls the optical depth
of the system, with results very similar to the homogeneous case,
and the case where the optical depth of the system is controlled
by the blocking effect of the clumps and is then proportional
to the filling factor (see Witt \& Gordon 1996).
Since the diffuse ISM of the disk models used here
has a doubly exponential distribution, clumps close to the center of the galaxy
have higher opical depths than those at the periphery of the disk.

Conversely, the dust clumps in the Monte Carlo simulations
of Bianchi et al. (2000) are distributed mostly in a ring structure
(Clemens et al. 1988; see also Sodroski et al. 1997)
and have all the same optical depth.
In addition, the description of the clumpiness of the dusty ISM of the disk
given by Bianchi et al. (2000) corresponds to quite low global filling factors,
for the three values of the fraction of the total gas mass
attributed by the latter authors to the molecular component of the ISM.
Bianchi et al. and Misiriotis \& Bianchi (2002) have already discussed
the differences in the behavior of the attenuation for the disk,
which arise as a consequence of these two different parameterizations
of the clumpy ISM, as a function of the inclination of the galaxy.

A low filling factor of the clumps (equal to 0.06) and a density contrast of 20
seem to reproduce better the structured ISM of low surface brightness galaxies
like UGC 7321 (Matthews \& Wood 2001).
The different set of parameters describing the clumpiness of the dusty ISM
in the Milky Way (Witt \& Gordon 1996) and in the low rotational velocity
Sd galaxy UGC 7321 may reflect systematic changes in the turbulent velocities
supporting the gas layer of a disk galaxy as a function of rotational velocity
(Dalcanton, Yoachim, \& Bernstein 2004).

Finally, constant, large optical depths (whatever the wavelength) are assumed
by Tuffs et al. (2004) to characterize the clumpy distribution of dust
in their models, that is associated with the opaque parent molecular clouds
of massive stars (see also Silva et al. 1998).
The attenuation produced by this clumpy component of the ISM
does not depend on the inclination of the galaxy (Tuffs et al. 2004).
Its wavelength dependence is due to the different blocking action of the clumps
on the photons emitted by stars of different masses (and, thus, ages).
Stars of different masses survive for different times,
so that lower mass, redder stars can escape further away
from the star-forming complexes during their longer lifetimes.

The models under investigation here include only Milky Way-type dust
(Witt \& Gordon 2000), while Ferrara et al. (1999) consider also
Small Magellanic Cloud-type dust (Witt \& Gordon 2000).
The values of albedo and, especially, asymmetry of the scattering
phase function for Milky Way-type dust given by Witt \& Gordon
and listed in Table 1 differ sensitively from those adopted
by Tuffs et al. (2004).

The doubly exponential disk models considered here
assume that the stars emitting their bulk luminosity at a given wavelength
have a larger scale height the longer the wavelength of the tracing photons
(see Table 1), but the same scale length whatever the wavelength.
This is at variance with the doubly exponential disk models
of Ferrara et al. (1999) and Matthews \& Wood (2001),
where the dust-to-stars scale-height ratio is independent of the wavelength.
Conversely, Tuffs et al. (2004) assume that
the old and young stellar populations of the disk are distributed
with two different scale heights but with the same scale length,
the latter increasing the shorter the wavelength.
In addition, in their disk models old and young stars
have different dust-to-stars scale-height ratios, independent of $\lambda$.

These differences reflect the present ambiguous knowledge
of the global structure of disks.
Structural studies of edge-on disk galaxies seem to reveal the ubiquity
of thick disks with red, old stellar populations similar to the thick disk
of the Milky Way (e.g. de Grijs \& van der Kruit 1996;
Dalcanton \& Bernstein 2002), but the magnitude of the intrinsic vertical
color gradients is uncertain (e.g. de Grijs \& Peletier 2000).
In addition, it is not established yet if the wavelength dependence
of the scale lengths of a disk galaxy represents also the existence
of stellar population gradients along the radial direction
(e.g. de Jong 1996; Kuchinski et al. 1998; Xilouris et al. 1999)
or is only due to dust attenuation (e.g. de Grijs 1998).

Finally, the bulge models under investigation here
have exponential radial profiles against the de Vaucouleurs' one
included by Ferrara et al. (1999) and Tuffs et al. (2004).
In all three cases the typical scale of the bulge does not depend
on the wavelength however.

In the previous sections, we have illustrated how the attenuation function,
the color excess, and the fraction of light scattered
towards the outside observer behave as a function
of the total central face-on extinction optical-depth at V band
(i.e., the opacity, equal to $2 \times \tau_V$ in our models) of the disk,
the inclination of the galaxy, and the structure (two-phase clumpy
vs. homogeneous) of the dusty ISM of the disk.

As a first result, we confirm that {\it the attenuation function
for either bulge or disk changes dramatically as a function
of both the disk opacity} (see Sect. 3.1.1),
regulating the total amount of dust of the models,
{\it and the inclination $i$ of the galaxy, for fixed dust/stars configuration
and structure of the dusty ISM of the disk} (see Sect. 3.2.1).
This has been already described by Ferrara et al. (1999),
Pierini et al. (2003), and Tuffs et al. (2004).
It is no surprise: it is a direct consequence of the definition
of the attenuation function as the combination of the extinction curve
with the geometry of a dusty stellar system.
The extinction curve describes
the combined absorption and out-of-the-beam scattering properties
of a mixture of dust grains of given size distribution and chemical composition
in a screen geometry as a function of wavelength.
It is an intrinsic property of the system, together with the total amount
of dust, the 3-D distribution of dust and stars, and the structure
of the dusty ISM.
Conversely, geometry depends also on the view angle of the system,
since the aspect of the dust/stars configuration changes
as a function of the observer's line of sight.
For all these reasons the behavior of the attenuation
as a function of the opacity and the inclination of a bulge$+$disk galaxy
is expected to be different for bulge and disk.
As a consequence, we find, not surprisingly (see e.g. Calzetti 2001),
that the attenuation function for either bulge or disk is very different
from the extinction function expected for a homogeneous, non-scattering
MW-type dust screen foreground to the light source.
The same applies to any color excess.

For the bulge of a face-on late-type galaxy,
the attenuation at a fixed wavelength $\lambda$ increases
with increasing $\tau_V$, while the attenuation function
becomes increasingly flat (or ``gray'') at those wavelengths
where the line-of-sight optical depth is larger than unity
across most of the region of the dust disk corresponding to
the face-on projection of the bulge (Fig. 1 and 2).
The structure of the dusty ISM induces larger differences
in the absolute value of the attenuation at a fixed $\lambda$
for intermediate values of $\tau_V$, but it does not affect
the shape of the attenuation function for $\tau_V \ge 2$.
In an analogous way, the attenuation function for a bulge
associated with a dust disk having fixed opacity
(e.g. the case of $\tau_V = 1$) becomes increasingly ``gray''
and less affected by the structure of the dusty medium when $i$ increases.
Nevertheless, a bulge suffers less attenuation
when the inclination of the galaxy approaches $\rm 90^o$,
since a larger fraction of its light is not intercepted by the dust disk.
This, of course, depends on the bulge-to-dust disk scale-height ratio,
as discussed by Ferrara et al. (1999).

For the disk of a face-on galaxy, the attenuation at a fixed $\lambda$
increases when $\tau_V$ increases, but the attenuation function
does not become as gray as for the face-on bulge.
In fact the {\it local} optical depths may still be low
in the outer regions of the disk.
Furthermore, for the face-on disk, the absolute values and shape
of the attenuation function depend more on $\tau_V$
than on the structure of the dusty ISM (Fig. 3 and 4).
Also for a disk with fixed opacity (e.g. the case of $\tau_V = 1$),
the absolute values and shape of the attenuation function
depend more on the inclination than on the structure of the dusty medium,
especially for $i \rm \ge 70^o$ (Fig. 15 and 16),
consistent with Matthews \& Wood (2001).
This is at variance with the result of Bianchi et al. (2000) however.
The reason for this discrepancy is the different description
of the distribution and the clumpiness of the dusty ISM,
as discussed by the latter authors.

The previous results on the attenuation function allow us to understand
the behavior of the color excess.
For a face-on bulge, we find that $E(B-V)$ reaches an asymptotic value
of $\sim 0.6$ mag already for $\tau_V \sim 2$, while $E(J-K)$ increases
up to 0.2--0.3 mag along the whole range in disk opacity probed by the models.
We remember that saturation of a color excess is reached
when line-of-sight optical depths of the order of unity or more are reached
for the effective wavelengths of both bands defining that color.
Conversely, for the face-on disk, both $E(B-V)$ and $E(J-K)$ increase
monotonically up to $\sim 0.1$ mag when $\tau_V$ increases.
The view angle of the galaxy affects the color excess, of course.
For a bulge associated with a disk having $\tau_V = 1$,
$E(B-V)$ stays constant up to $i \rm \sim 60^o$ and drops almost by half
for higher inclinations, while $E(J-K)$ increases from $\sim 0.05$
up to $\sim 0.2$ mag, when $i$ reaches $\rm 80^o$ and drops to 0.14 mag
for $i \rm = 90^o$.
Conversely, for a disk with $\tau_V = 1$, both $E(B-V)$ and $E(J-K)$ increase
monotonically when $i$ increases.
In general, a homogeneous dust distribution produces larger values
of $E(B-V)$ and $E(J-K)$.
However, for systems seen almost edge on, the structure of the dusty ISM
has no impact on the color excess for either bulge or disk,
consistent with the result of Matthews \& Wood (2001)
for their edge-on disk models.
In general, the structure of the dusty ISM of the disk
has a larger effect on the observed magnitudes than on the observed colors
of either bulge or disk component, according to our models.

The study of the behavior of the optical and near-IR color excesses
of bulge and disk, as a function of disk opacity and galaxy inclination
has important consequences for the interpretation of the colors
of galaxies spanning the whole sequence of late Hubble types.
For example, Fioc \& Rocca-Volmerange (1999) find that
the observed total $\rm J-H$ and $\rm H-K$ colors of Sa--Sbc galaxies
are redder than those of elliptical and lenticular galaxies
by 0.04 and 0.05 mag, respectively.
They comment that ``the differential internal extinction,
negligible in the near-IR, is also unlikely to cause this phenomenon''.
Our previous results clearly contradict their conclusion.
{\it These results hint at a sensitive differential effect
of the internal extinction in the near-IR,
especially for the (exponential) bulge of a late-type galaxy}.
In case of a galaxy with average inclination $i \rm \sim 60^o$
and $\tau_V \sim 1$, an exponential bulge may by reddened
by $E(J-K) \sim 0.1$ mag (Fig. 19) against a value of $E(J-K) \sim 0.04$ mag
for its disk (Fig. 21).
These figures are consistent with the fact that observed,
total $\rm J-K$ colors of bulge-dominated, early-type spiral galaxies
are redder than those of E/S0 galaxies by 0.09 mag, when taking into account
that these galaxies have more $r^{-1/4}$-like bulges
(see e.g. Andredakis et al. 1995; Graham 2001; Hunt et al. 2004).

Furthermore, we find that
{\it a measure of the total-to-selective attenuation ratio
is not equivalent to a measure of the total-to-selective extinction ratio},
that characterizes the dust mixture present in a dusty galaxy:
the former ratio changes a lot as a function of the opacity and inclination
of the system.

This study focuses on the properties of the scattered light more than others,
also including anisotropic multiple scattering, for bulge$+$disk systems.
We find, not surprisingly (e.g. Witt \& Gordon 1996), that
a homogeneous dust distribution absorbs and scatters more light
at any fixed $\lambda$ than a two-phase clumpy one,
whether this light was produced either in the bulge or in the disk,
for almost all values of $\tau_V$ and $i$.
In fact, for the edge-on view of a spiral galaxy with $\tau_V = 1$,
the fraction of light emitted by either the bulge or the disk
towards the outside observer does not depend on the clumpiness
of the dusty ISM, whatever $\lambda$.

{\it The fraction of light produced either in the bulge or in the disk
at a fixed $\lambda$ and scattered towards the outside observer
is also a function of $\tau_V$ and $i$,
and so is the color of the total scattered light}.
In particular, for a given $\lambda$, this fraction increases
with increasing disk-averaged optical depth
along the observer's line of sight, up to the point that
most of the single and multiple scattered photons get eventually absorbed
or leave the system in a direction different from the observer's line of sight
(especially for optical/near-IR wavelengths).
Dust grains scatter more efficiently and in a more forward direction,
as well as absorb, the shorter the wavelength of the photons (see Table 1).
Thus, for a fixed inclination (e.g. $i \rm = 0^o$),
the color of the light emitted from either the bulge or the disk
and scattered towards the outside observer will turn from blue to red
when $\tau_V$ increases.
Conversely, for a fixed value of disk opacity
(e.g. in the case of $\tau_V = 1$), the color of the light emitted
by either bulge or disk and scattered towards the outside observer
will stay almost the same when $i$ increases (except for the edge-on disk).

{\it For the light received by the outside observer and produced
either in the bulge or in the disk at a given wavelength,
the fraction which is scattered by the dust corresponds
up to 20--30 per cent of the fraction which reaches the observer
without being affected by the dust, for the face-on bulge,
and up to 10--95 per cent for the face-on disk,
as a function of the opacity and the structure of the dusty ISM}.
For the outside observer, the ratio of scattered-to-directly transmitted light,
$f(sca)/f(dir)$, decreases (increases) with increasing $\tau_V$
for the face-on bulge (disk), whatever the emission wavelength
and whatever the structure of the dusty medium.
This ratio decreases with increasing inclination for the bulge of a galaxy
with $\tau_V$ equal to unity, whatever $\lambda$
and whatever the structure of the dusty ISM.
For the disk of the same galaxy, the behavior of $f(sca)/f(dir)$
as a function of $i$ is more complex.
This time $f(sca)/f(dir)$ increases with increasing inclination,
for $\rm 0^o \le$ $i$ $\rm \le 70^o$, whatever $\lambda$
and whatever the structure of the dusty ISM;
it decreases with increasing inclination, for $\rm 70^o <$ $i$ $\rm \le 90^o$,
when $\lambda \rm \ge 0.2~\mu m$, whatever the structure of the dusty ISM;
it increases (stays almost constant) with increasing inclination,
for $\rm 70^o <$ $i$ $\rm \le 90^o$, when $\lambda \rm < 0.2~\mu m$
and the structure of the dusty ISM is two-phase clumpy (homogeneous).
Of course this result depends also on the wavelength dependence
of the albedo and the scattering phase function asymmetry parameter
assumed for the mixture of dust grains and on the relative distribution
of photons of given wavelength and dust within the disk (see Table 1).
In general, the contribution of the scattered radiation
to the total monochromatic light received by the outside observer
is significant, especially at UV wavelengths,
even for a two-phase, clumpy, dusty ISM.
Hence understanding dust scattering properties
(see e.g. Gordon 2004 for a discussion) is fundamental for the interpretation
of e.g. the UV extragalactic observations made with {\it GALEX}
(Milliard et al. 2001).

Finally, our investigation confirms that {\it radiative transfer effects
may dilute the relative prominence of the extinction feature}
at $\rm 2175~\AA$, distinctive of the Milky Way-type dust
present in our models, when the attenuation optical depth in the far-UV range
is large enough, for either bulge or disk, whatever the structure
of the dusty ISM.
This is in agreement with the results of Cimatti et al. (1997)
and Ferrara et al. (1999).
The latter authors attributed the disappearence of this feature
to scattering modulation due to geometry and multiple scattering of photons
on dust grains.
In fact the probability for events of (multiple) scattering increases
when the disk-averaged optical depth along the observer's line of sight
increases.
However, multiple scattered photons emitted at neighboring wavelengths
of the $\rm 2175~\AA$-extinction bump will be eventually absorbed,
owing to the wavelength-dependence of the scattering phase function
asymmetry parameter.
Hence absorption will increase faster at neighboring wavelengths
of $\rm 2175~\AA$.

The models of a bulge$+$disk galaxy used here
simulate structural characteristics of an Sbc galaxy like our own,
nevertheless the results of Sect. 3.1.2 and 3.2.2
may be qualitatively applied to spiral galaxies of earlier type
(cf. Ferrara et al. 1999 and Tuffs et al. 2004).
For a quantitative conclusion, further modeling is needed,
which takes into account observed scaling relations
for bulge and disk components as a function of Hubble type and luminosity
(e.g. Andredakis et al. 1995; Graham 2001; Hunt et al. 2004)
as well as the observed dependence of disk opacity on galaxy luminosity
(Giovanelli et al. 1995; Wang \& Heckman 1996; Tully et al. 1998;
Masters et al. 2003).
These physical aspects are not easy to single out from observations, however,
as discussed by Tuffs et al. (2004).
In addition, sytematic differences in the dust/stars configurations
and in the structure of the dusty ISM seem to exist
between giant and dwarf late-type galaxies (Dalcanton et al. 2004)
and need to be taken into account.
We will address all these issues in future publications of this series.

\acknowledgments

This work was supported partly through grant NAG 5-9202
from the National Aeronautics and Space Administration
to the University of Toledo.
We are grateful to the referee, W. C. Keel, for providing
perceptive and helpful comments.



\clearpage

\begin{figure}
\epsscale{.80}
\plotone{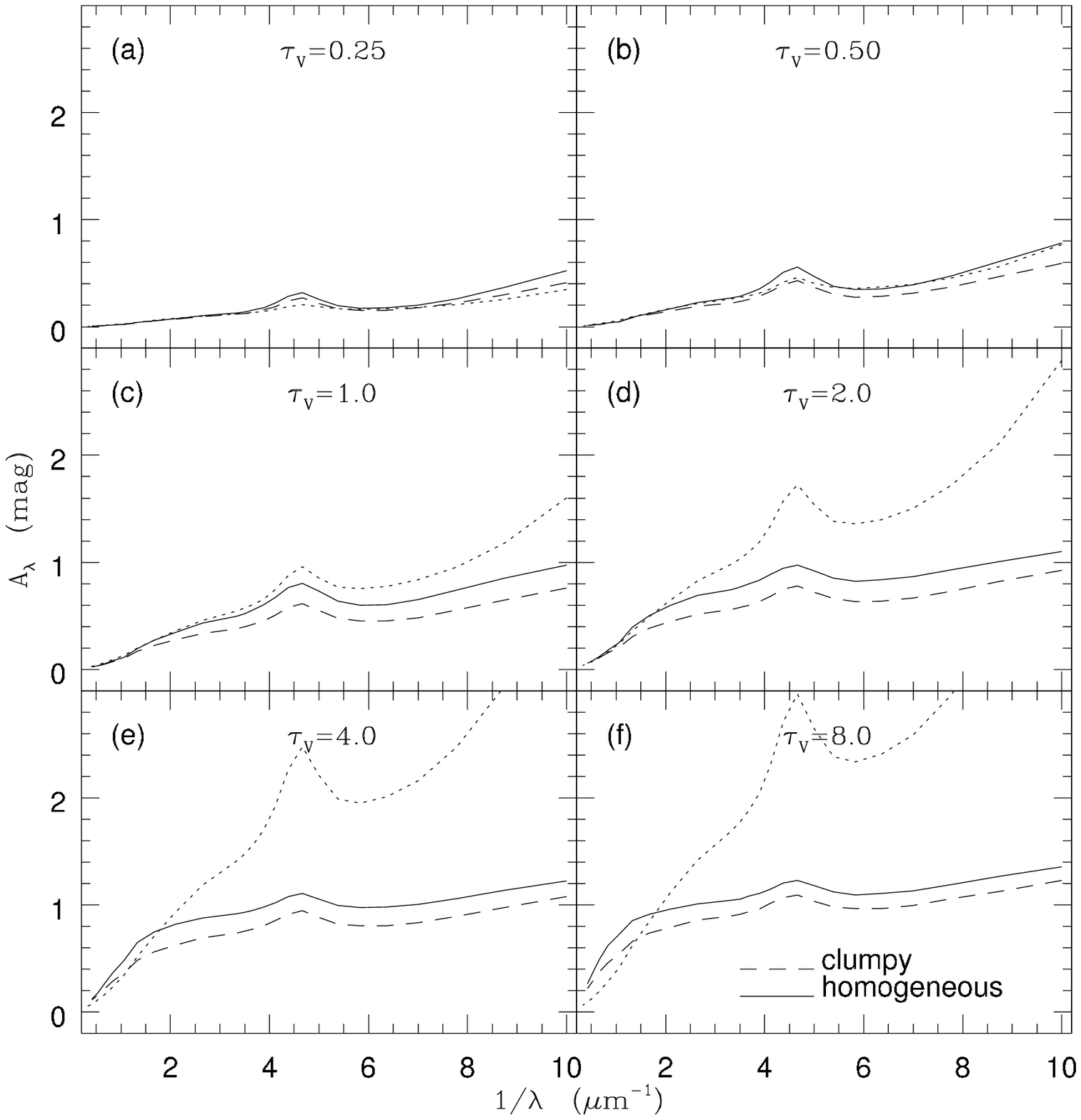}
\caption{The attenuation function $A_{\lambda}$ for the bulge
of a late-type galaxy seen face on as a function of the opacity
(equal to $2 \times \tau_V$ in our models) and the structure of the dusty ISM
of the disk. Hereafter models with a two-phase clumpy (homogeneous)
dust distribution are reproduced with short-dashed (solid) lines.
Different panels illustrate $A_{\lambda}$ for models with different values
of $\tau_V$. It is evident, in particular, that a homogeneous dust distribution
provides a larger attenuation than a two-phase clumpy one, for any fixed value
of the disk opacity. This gap is maximal (a factor of $\sim$ 1.6)
for $\tau_V \sim 2$. In addition, we draw the extinction function expected
for a homogeneous, non-scattering MW-type dust screen foreground
to the light source (dotted line) with total V-band extinction optical depth
equal to the total attenuation optical depth at V band of the face-on
bulge model with a homogeneous dust disk reproduced in each panel.
The comparison confirms that the dust obscuration for an extended object
like a bulge is conceptually different from the dust extinction of stars.
\label{fig1}}
\end{figure}

\clearpage

\begin{figure}
\epsscale{.80}
\plotone{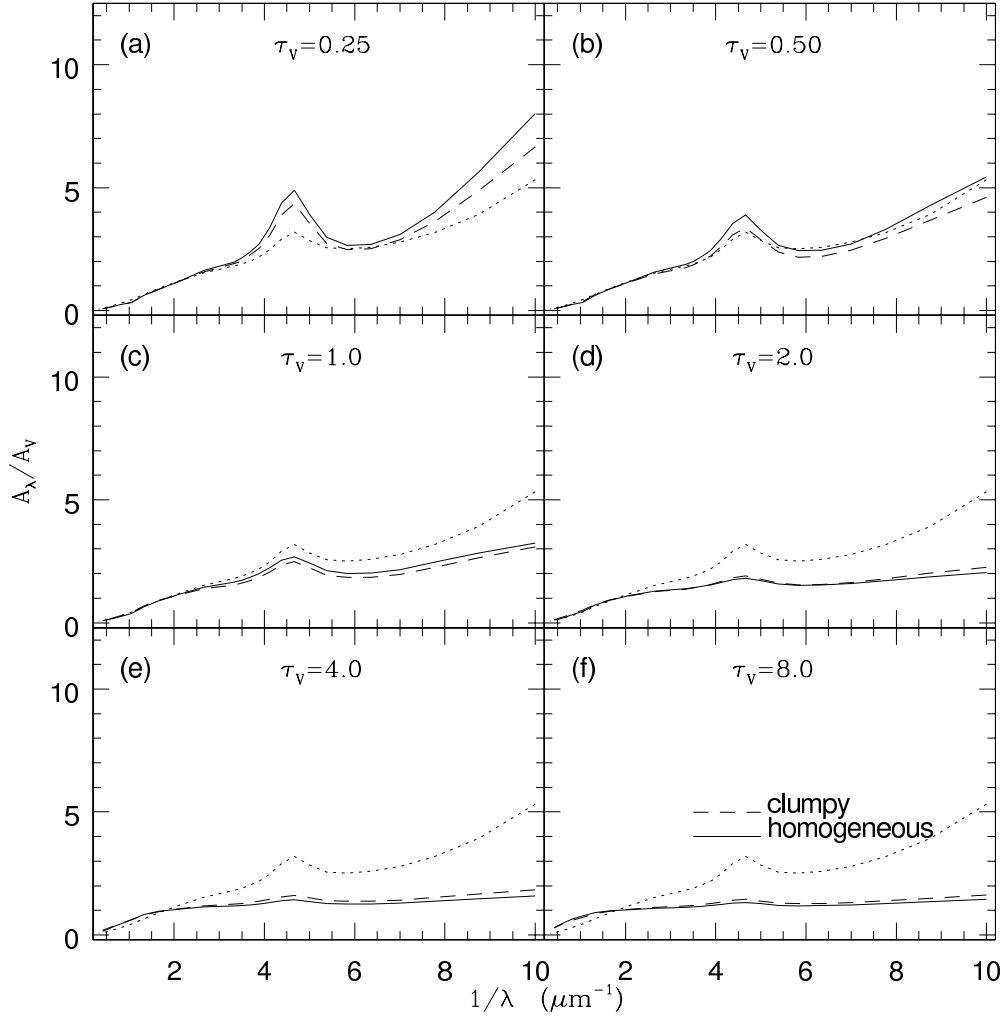}
\caption{The attenuation function normalized to its V-band value,
$A_{\lambda}/A_V$, for a face-on bulge as a function of the opacity
and the structure of the dusty ISM of the associated disk. Different panels
illustrate $A_{\lambda}/A_V$ for models with different values of $\tau_V$.
It is evident that the shape of the attenuation function for a face-on bulge
does not depend much on the local dust distribution (homogeneous vs. two-phase
clumpy). In addition, we note that the relative prominence of the absorption
feature at $\rm \sim 2175~\AA$ and the relative steepness of $A_{\lambda}/A_V$
in the far-UV vanish towards larger values of $\tau_V$. Finally, in each panel
we draw the extinction function normalized to its V-band value
for a homogeneous, non-scattering MW-type dust screen foreground
to the light source (dotted line), whatever the total V-band extinction
optical depth. This normalized extinction function is equal to
the MW extinction curve normalized to its V-band value listed in Table 1.
\label{fig2}}
\end{figure}

\clearpage

\begin{figure}
\epsscale{.80}
\plotone{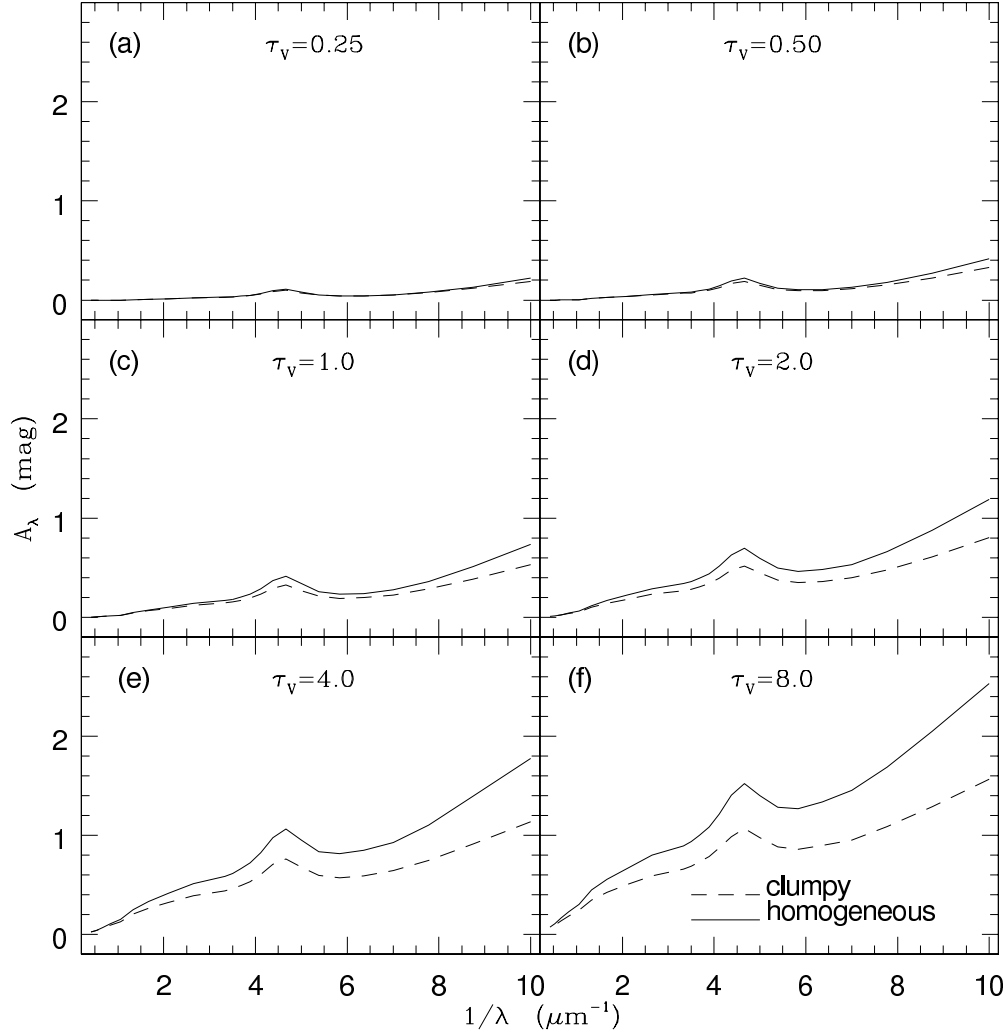}
\caption{The attenuation function for the disk of a late-type
galaxy seen face on as a function of the opacity and the local distribution
of the dust. Different panels illustrate $A_{\lambda}$ for models
with different values of $\tau_V$. For a face-on disk, $A_{\lambda}$ increases
when $\tau_V$ increases, for $\tau_V \le 8$. The slope of the attenuation
function increases as well. As a consequence, $A_{\lambda}$ does not reach
any asymptotic value.\label{fig3}}
\end{figure}

\clearpage

\begin{figure}
\epsscale{.80}
\plotone{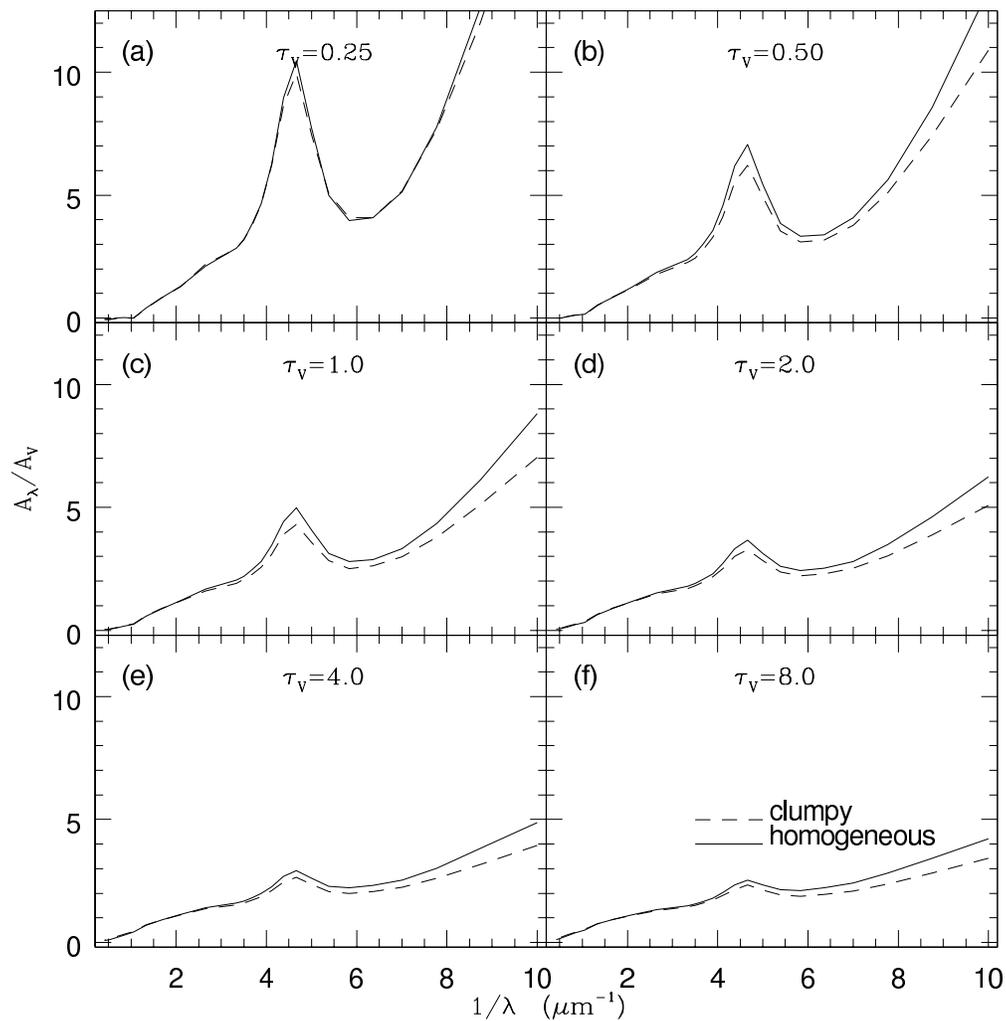}
\caption{$A_{\lambda}/A_V$ for a face-on disk as a function of
$\tau_V$ and the structure of its dusty ISM. Different panels illustrate
$A_{\lambda}/A_V$ for models with different values of $\tau_V$. In general,
analogous considerations to a face-on bulge (Fig. 2) apply to a face-on disk,
but the $\rm 2175~\AA$-extinction bump and the relative steepness
of $A_{\lambda}/A_V$ in the far-UV are more evident for the latter.
\label{fig4}}
\end{figure}

\clearpage

\begin{figure}
\epsscale{.80}
\plotone{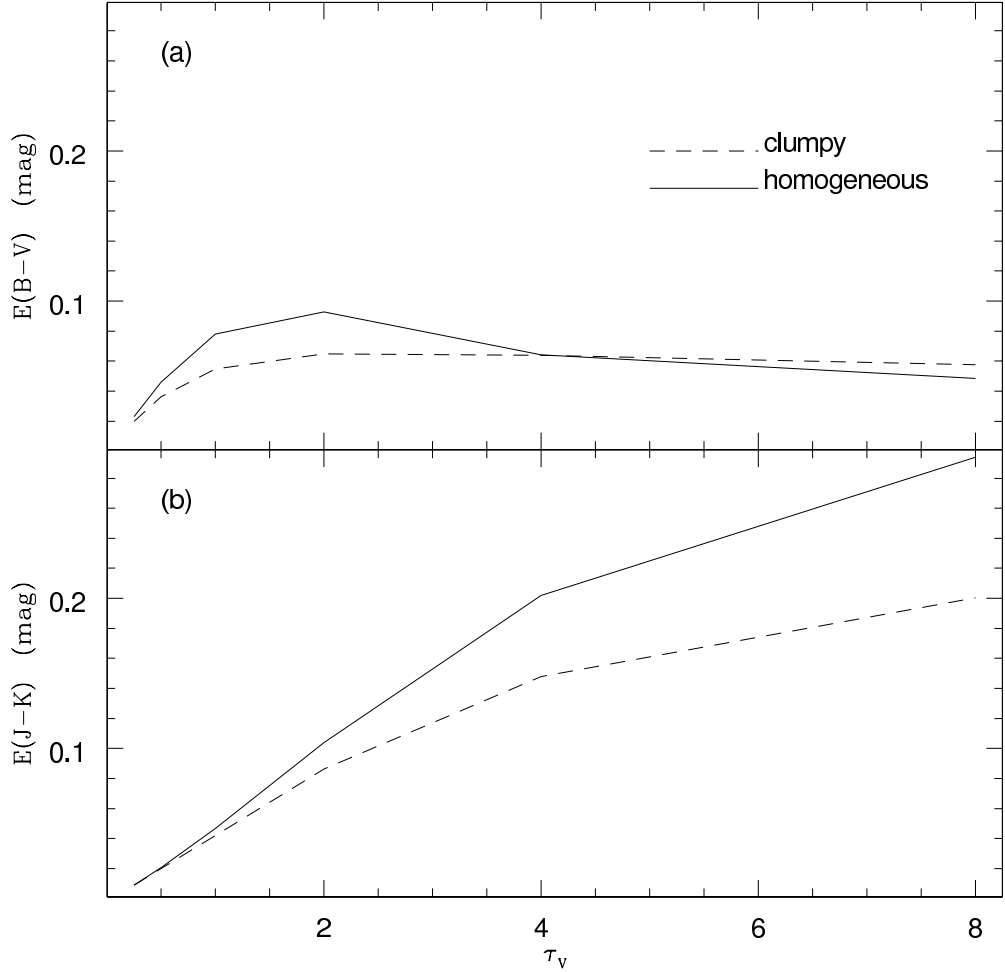}
\caption{Optical and near-IR color excesses, $E(B-V)$ (a)
and $E(J-K)$ (b), respectively, vs. $\tau_V$ for the face-on bulge,
as a function of the structure of the dusty ISM of the associatde disk.
$E(B-V)$ reaches a (flat) maximum when the line-of-sight optical depths
at 0.44 and $\rm 0.55~\mu m$ are close to unity in the region of the dust disk
corresponding to the face-on projection of the bulge. Conversely,
no saturation is reached by $E(J-K)$ for $\tau_V \le 8$, owing to
the much lower line-of-sight optical depths at 1.25 and $\rm 2.17~\mu m$,
whatever the local distribution of the dust in the disk.\label{fig5}}
\end{figure}

\clearpage

\begin{figure}
\epsscale{.80}
\plotone{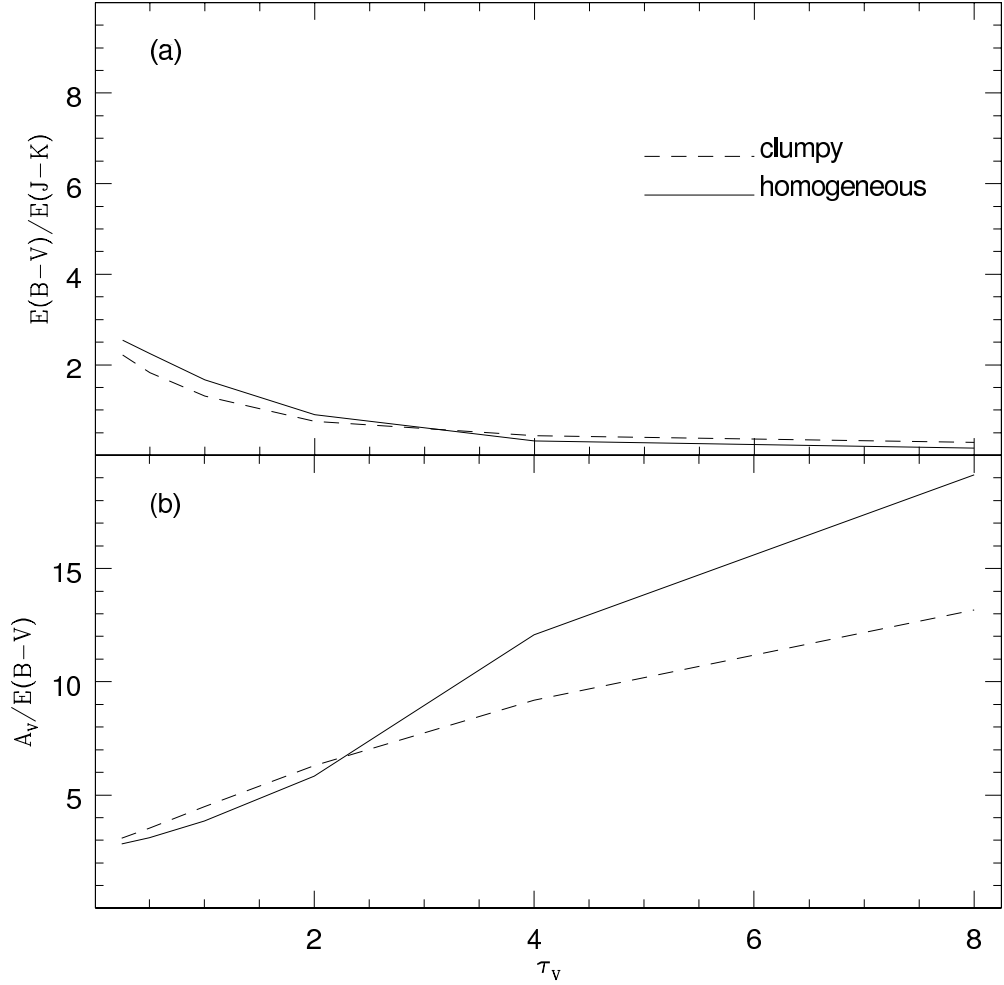}
\caption{$E(B-V)/E(J-K)$ (a) and $R_V = A_V/E(B-V)$ (b)
vs. $\tau_V$ for the face-on bulge, as a function of the structure
of the dusty ISM of the associated disk. The behavior of $E(B-V)/E(J-K)$
may be easily understood from Fig. 5. Interestingly, the total-to-selective
{\it attenuation} ratio $R_V$ increases non-linearly with increasing $\tau_V$,
starting from a value similar to 3.1, which is the total-to-selective
{\it extinction} ratio for the standard MW extinction curve.
$R_V$ strongly depends on the structure of the dusty ISM for $\tau_V > 2$.
\label{fig6}}
\end{figure}

\clearpage

\begin{figure}
\epsscale{.80}
\plotone{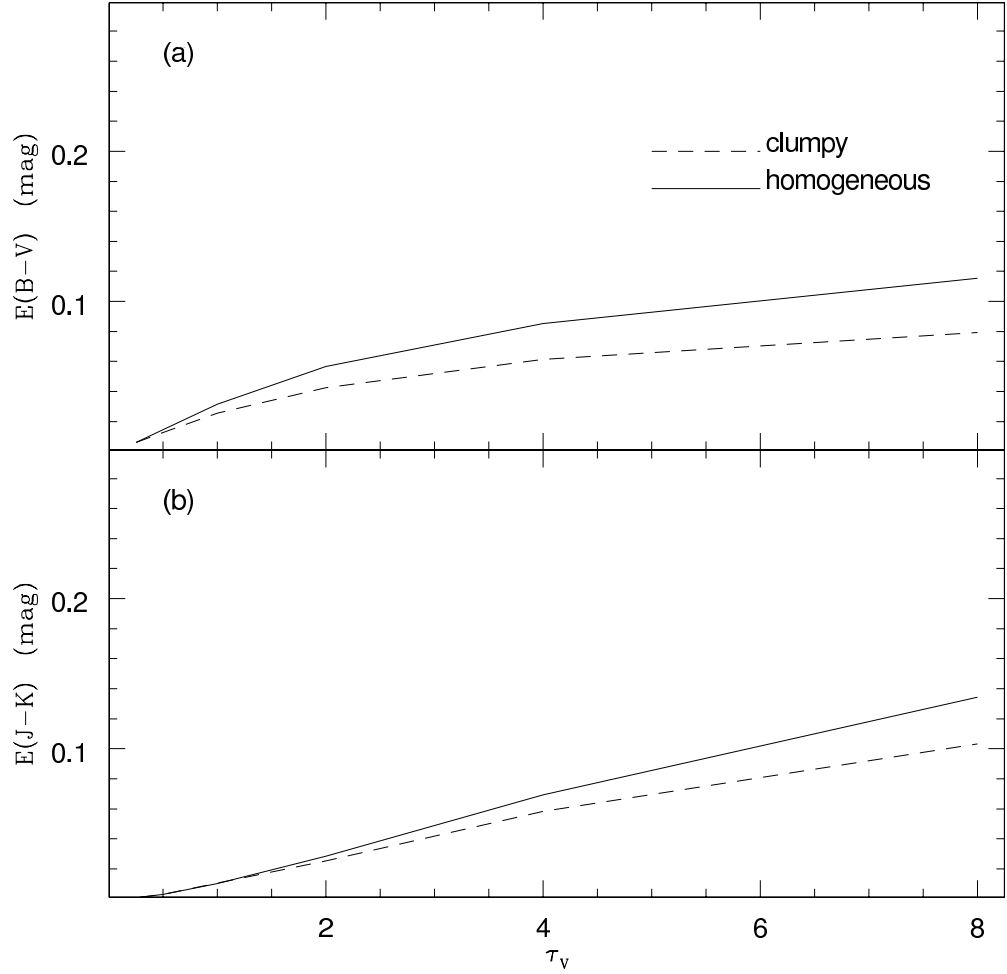}
\caption{$E(B-V)$ (a) and $E(J-K)$ (b) vs. $\tau_V$ for
the face-on disk, as a function of the structure of the dusty ISM.
No saturation is reached by $E(B-V)$ and $E(J-K)$ for $\tau_V \le 8$,
whatever the local distribution of the dust.\label{fig7}}
\end{figure}

\clearpage

\begin{figure}
\epsscale{.80}
\plotone{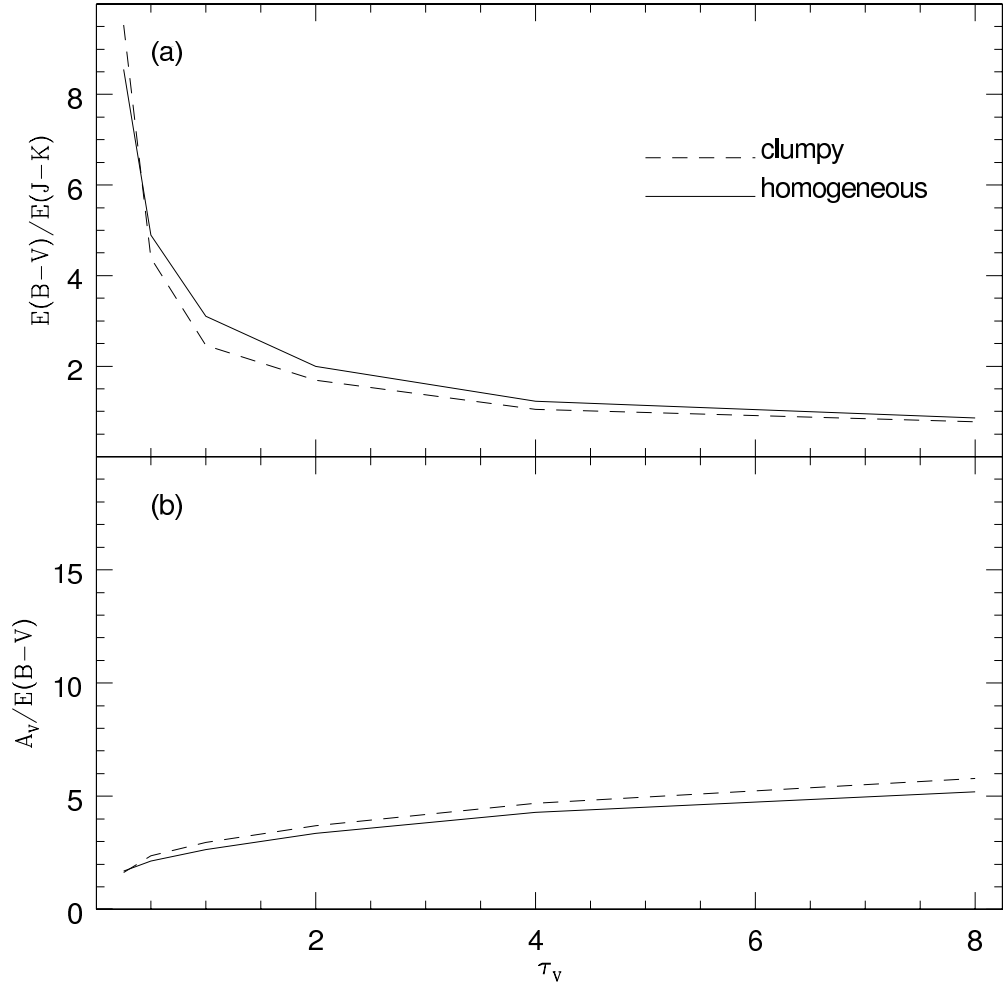}
\caption{$E(B-V)/E(J-K)$ (a) and $A_V/E(B-V)$ (b) vs. $\tau_V$
for the face-on disk, as a function of the structure of the dusty ISM.
Note the difference from the case of the face-on bulge (Fig. 6).\label{fig8}}
\end{figure}

\clearpage

\begin{figure}
\epsscale{.80}
\plotone{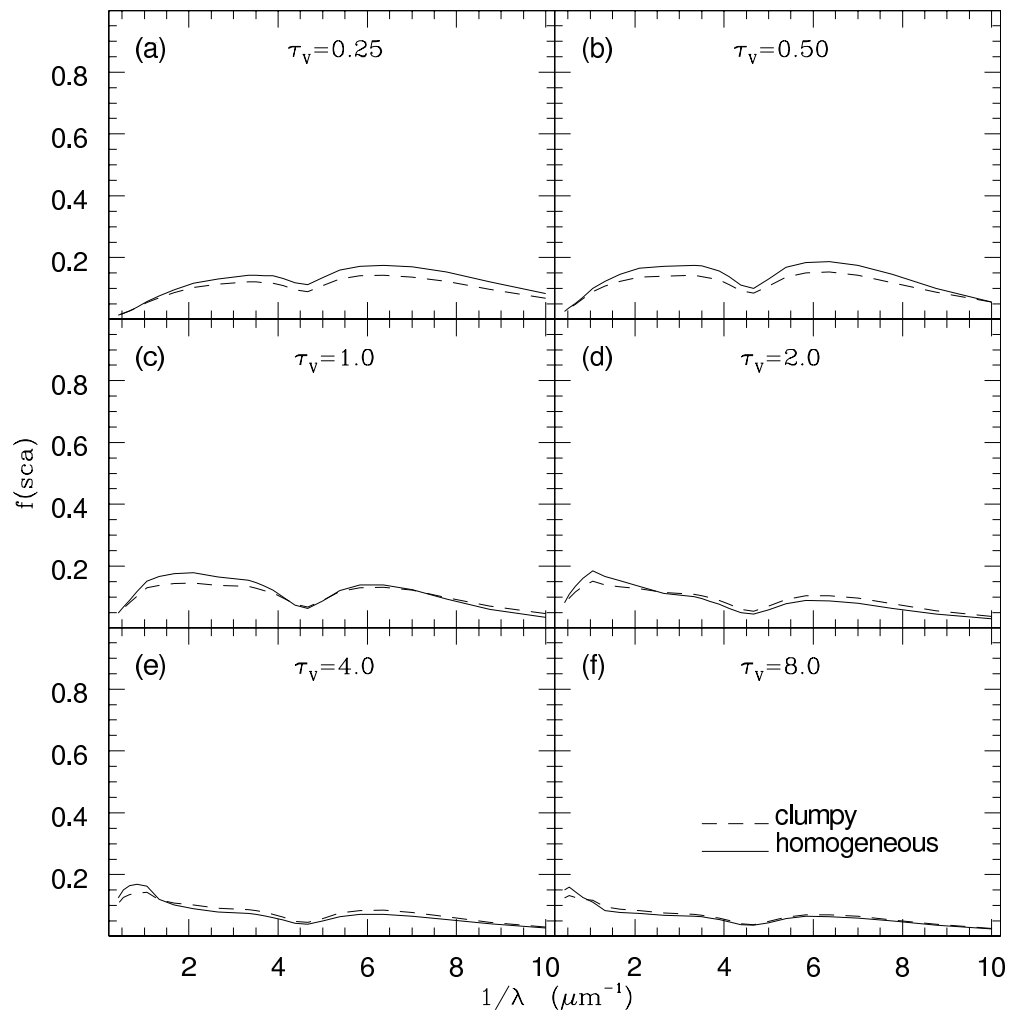}
\caption{The fraction of light which is emitted at $\lambda$
and is scattered towards the outside observer, $f(sca)$, as a function of
the opacity and the structure of the dusty ISM of the disk,
for the face-on bulge. Note that the color of the scattered light
turns from blue to red when $\tau_V$ increases, owing to the increase
of the effective albedo at longer wavelengths and the eventual absorption
of multiple-scattered far-UV photons.\label{fig9}}
\end{figure}

\clearpage

\begin{figure}
\epsscale{.80}
\plotone{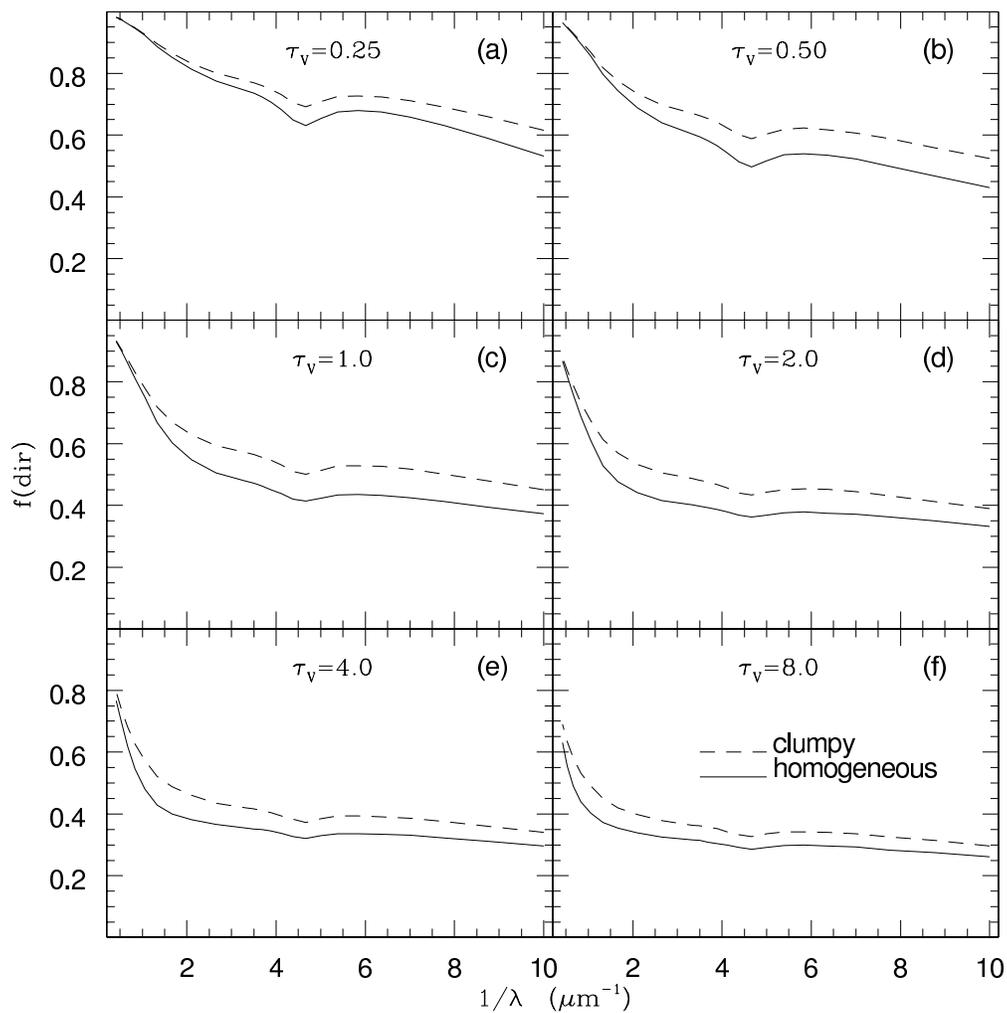}
\caption{The fraction of light which is emitted at $\lambda$
and is directly transmitted towards the outside observer, $f(dir)$,
as a function of the opacity and the structure of the dusty ISM of the disk,
for the face-on bulge. Note the flattening of $f(dir)$ at progressively larger
wavelengths, for $\tau_V \ge 2$, when effective optical depths
progressively larger than unity are reached through increasingly broader
regions of the dust disk, whatever the local distribution of the dust.
\label{fig10}}
\end{figure}

\clearpage

\begin{figure}
\epsscale{.80}
\plotone{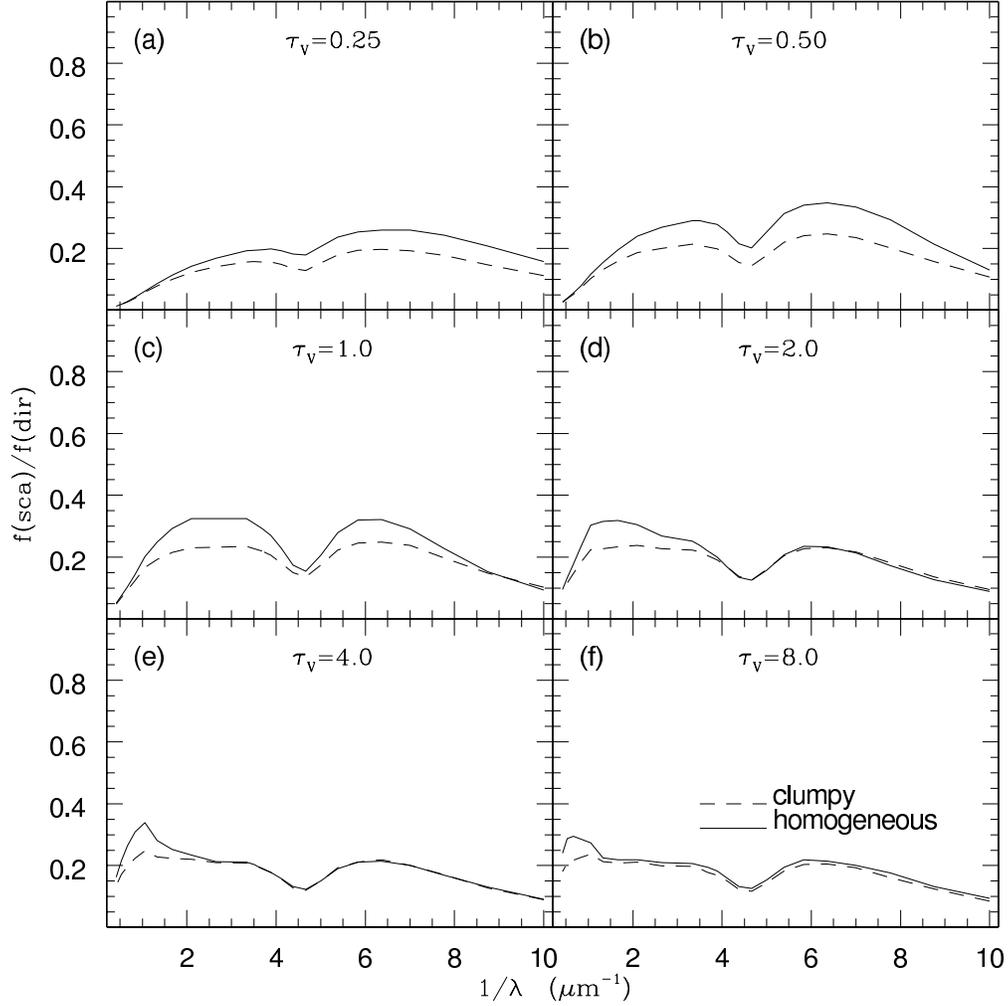}
\caption{The ratio of scattered-to-directly transmitted light,
$f(sca)/f(dir)$, for the radiation received by the outside observer
from the face-on bulge, as a function of the opacity and the structure
of the dusty ISM of the associated disk. $f(sca)/f(dir)$ has a maximum value
of the order of 30 per cent and its peak shifts towards longer wavelengths
with increasing opacity. In general, $f(sca)/f(dir)$ is lower
for a two-phase clumpy dust distribution than for a homogeneous one.
\label{fig11}}
\end{figure}

\clearpage

\begin{figure}
\epsscale{.80}
\plotone{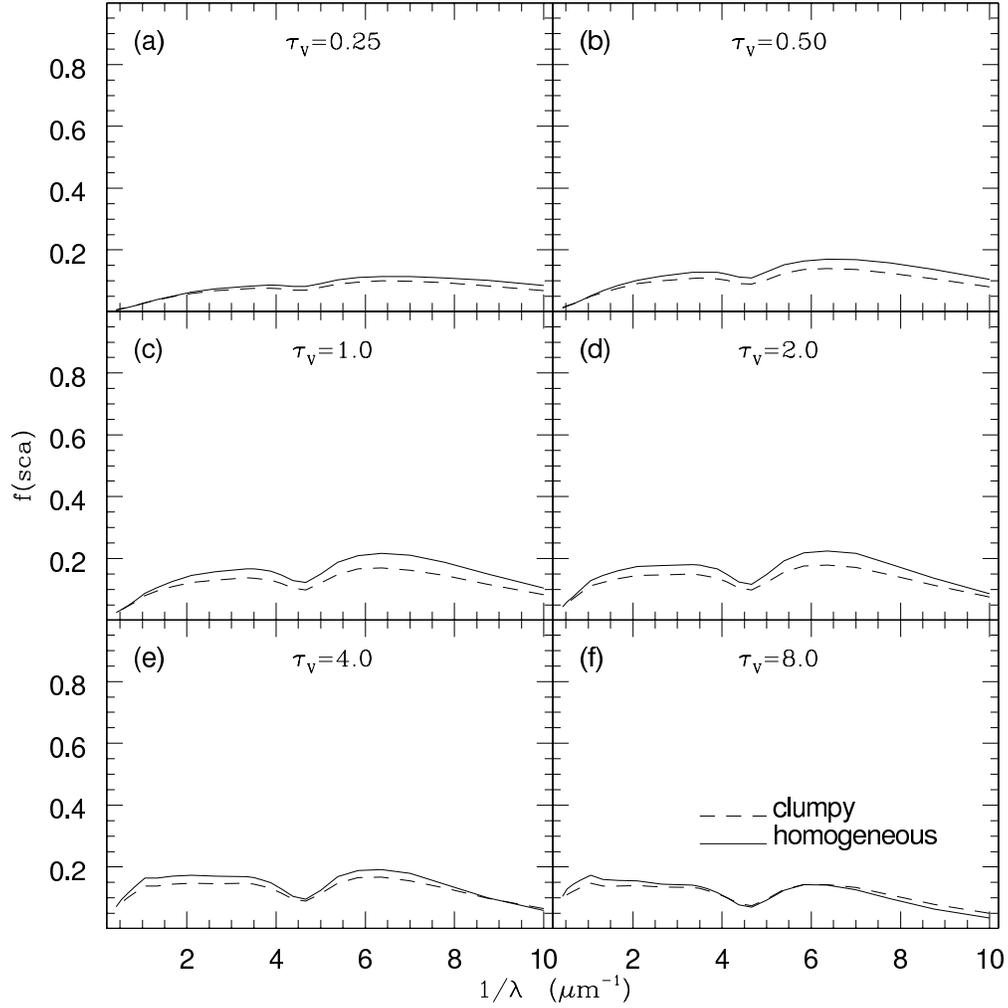}
\caption{$f(sca)$ as a function of the opacity
and the structure of the dusty ISM for the face-on disk. In this case,
a two-phase, clumpy, dusty ISM produces a reduced effective albedo
with respect to a homogeneous one at almost every $\lambda$, for $\tau_V \le 8$
(cf. instead Fig. 9).\label{fig12}}
\end{figure}

\clearpage

\begin{figure}
\epsscale{.80}
\plotone{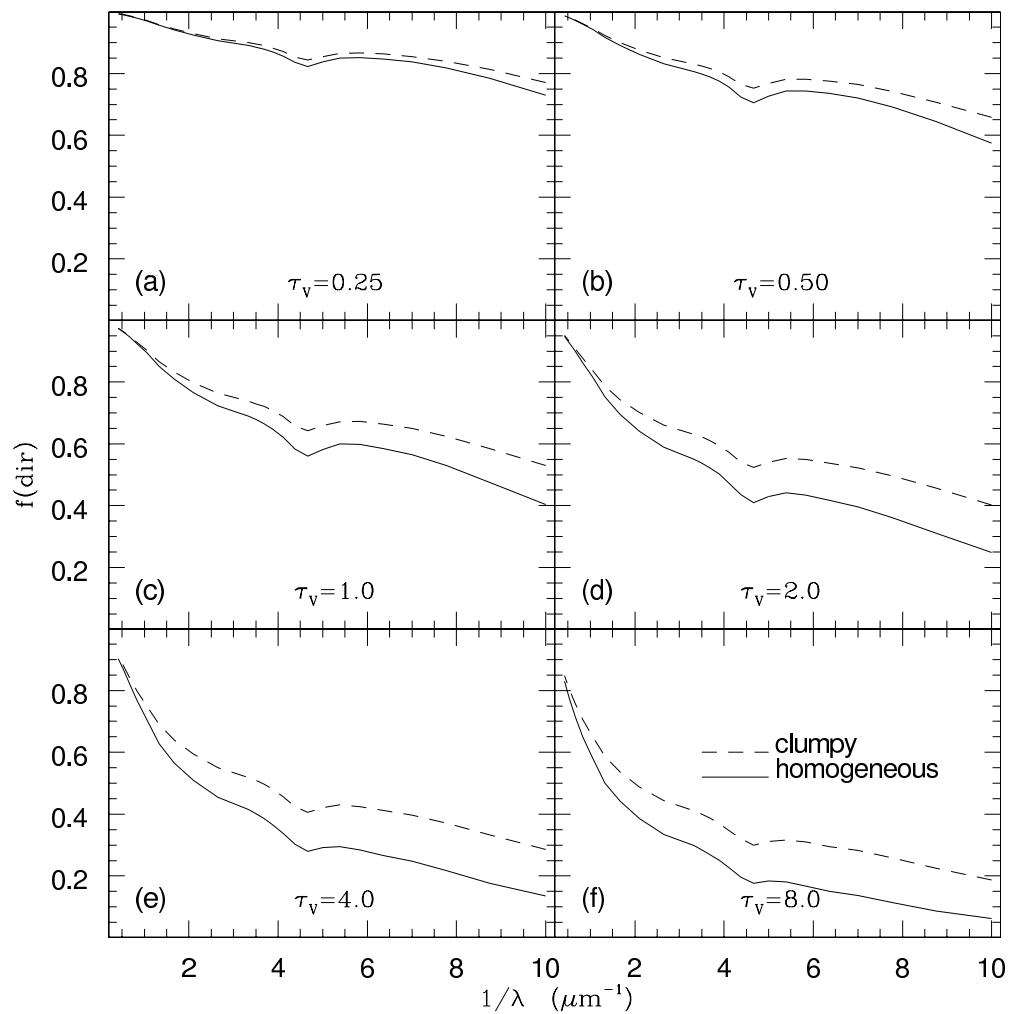}
\caption{$f(dir)$ as a function of the opacity
and the structure of the dusty ISM for the face-on disk. In this case,
$f(dir)$ is a monotonically decreasing function of $\lambda$ for $\tau_V \le 8$
(cf. instead Fig. 10).\label{fig13}}
\end{figure}

\clearpage

\begin{figure}
\epsscale{.80}
\plotone{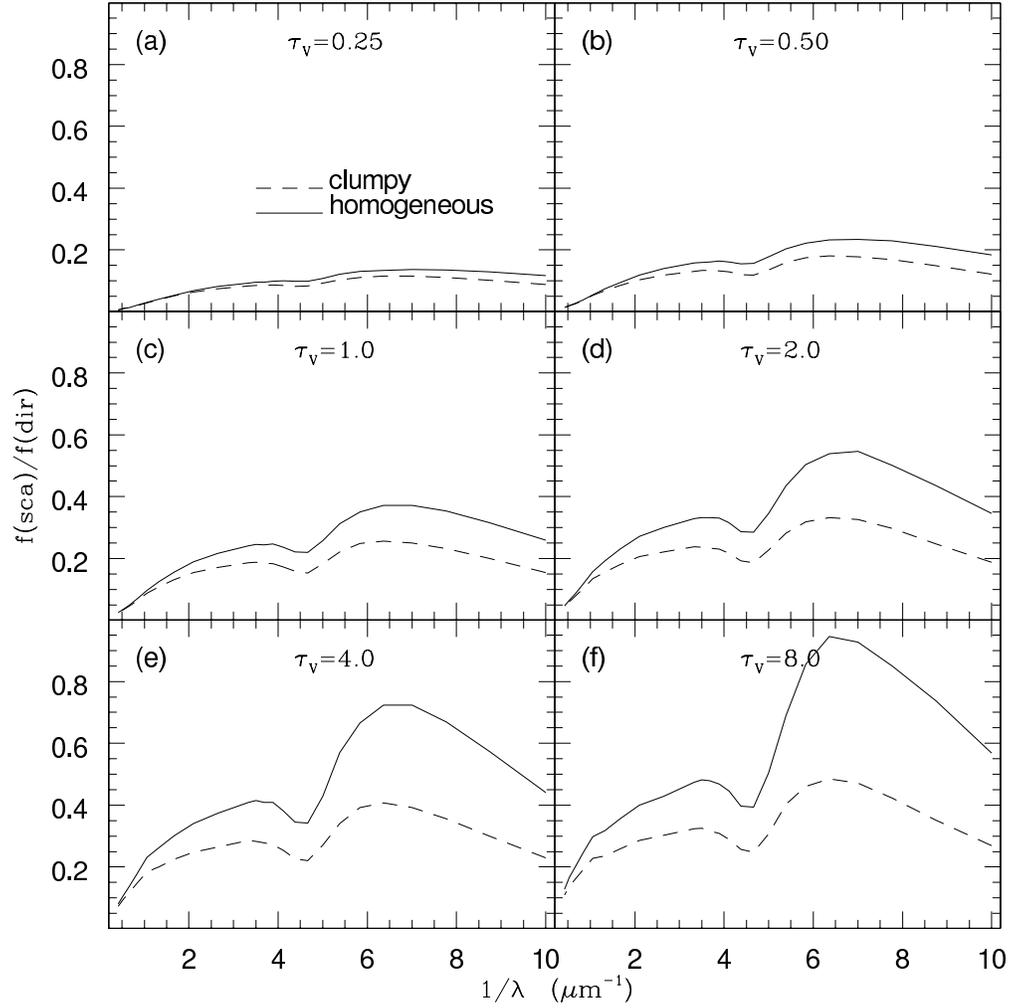}
\caption{$f(sca)/f(dir)$ as a function of the opacity
and the structure of the dusty ISM for the face-on disk. In this case,
$f(sca)/f(dir)$ increases with increasing opacity at any wavelength,
but always peaks at around $\rm 0.15~\mu m$ for $\tau_V \le 8$. This behavior
is opposite to that of the face-on bulge (cf. Fig. 11).\label{fig14}}
\end{figure}

\clearpage

\begin{figure}
\epsscale{.50}
\plotone{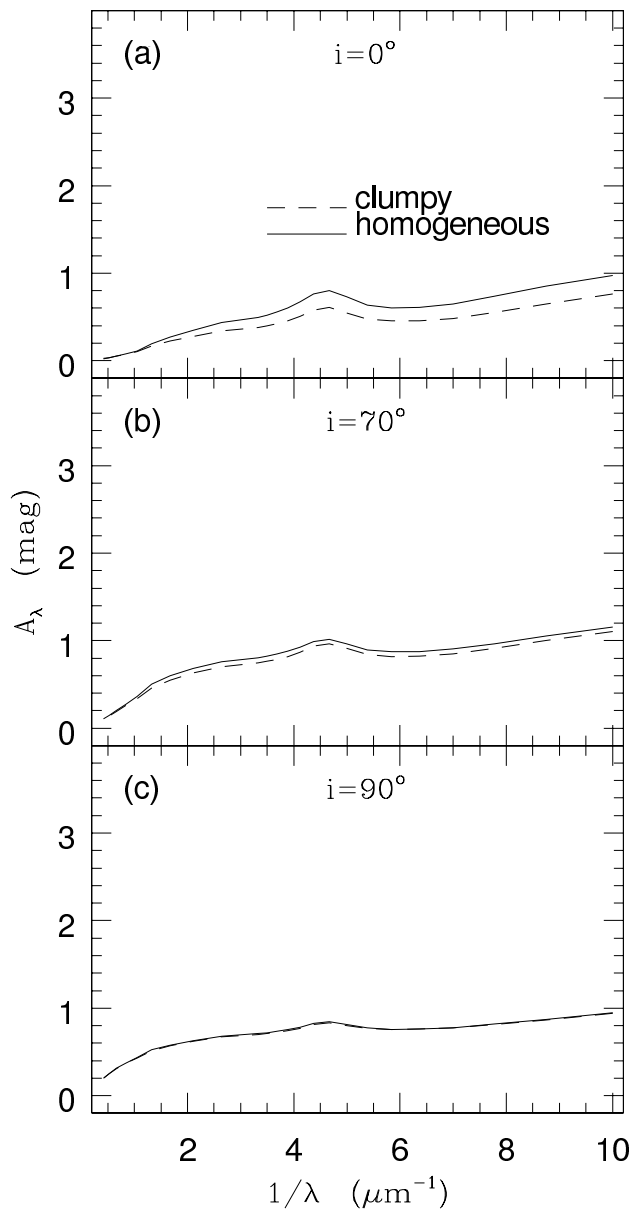}
\caption{The attenuation function of the bulge of a late-type galaxy
with $\tau_V = 1.0$, as a function of the inclination
and the structure of the dusty ISM of the disk. Three cases are reproduced:
$i \rm = 0^o$ (a), $\rm 70^o$ (b), and $\rm 90^o$ (c). $A_{\lambda}$ flattens
with increasing inclination, owing to the increase of the line-of-sight
optical depth across the region of the dust disk overlapping
with the projection of the bulge. The structure of the dusty ISM of the disk
plays no role for $i \rm \ge 70^o$.\label{fig15}}
\end{figure}

\clearpage

\begin{figure}
\epsscale{.50}
\plotone{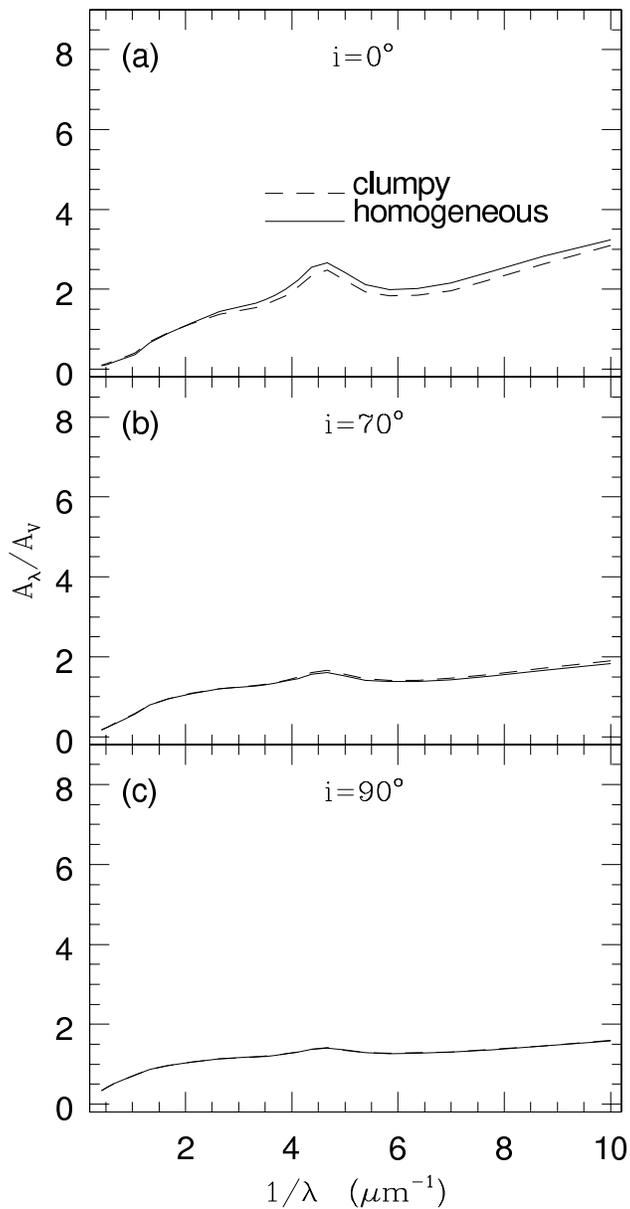}
\caption{The normalized attenuation function of the bulge
of a late-type galaxy with $\tau_V = 1.0$, as a function of the inclination
and the structure of the dusty ISM of the disk. Three cases are reproduced:
$i \rm = 0^o$ (a), $\rm 70^o$ (b), and $\rm 90^o$ (c).
$A_{\lambda}/A_V$ becomes ``grey'' with increasing inclination,
as expected from Fig. 15.\label{fig16}}
\end{figure}

\clearpage

\begin{figure}
\epsscale{.50}
\plotone{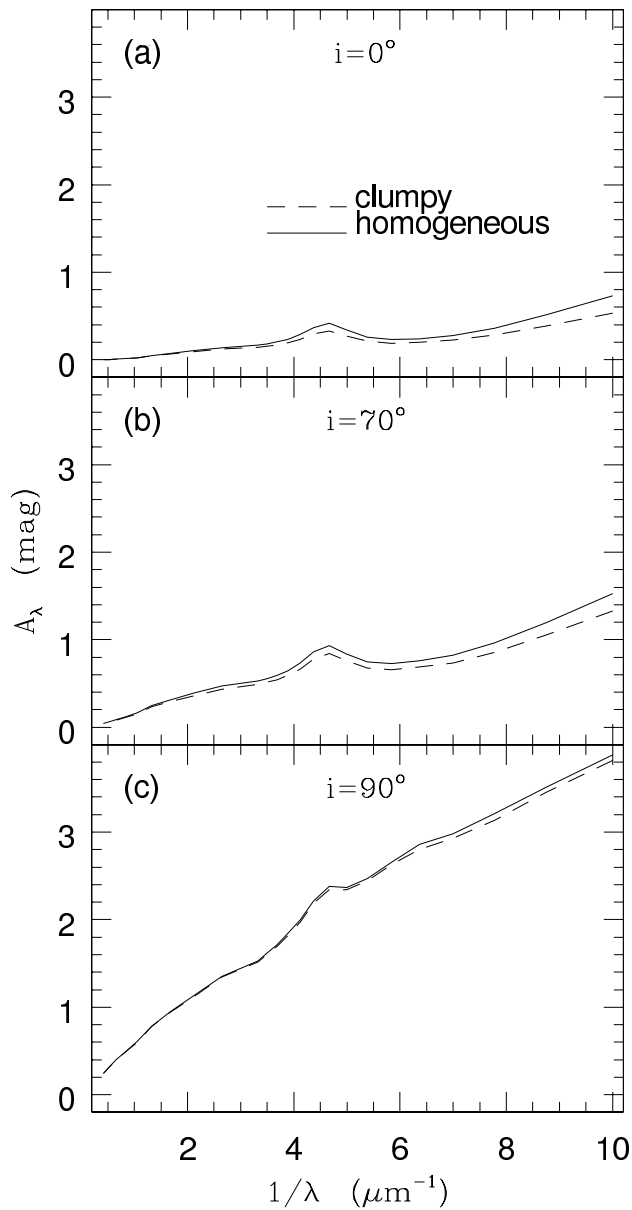}
\caption{The attenuation function of the disk of a galaxy
with $\tau_V = 1.0$, as a function of the inclination
and the structure of the dusty ISM. Three cases are reproduced:
$i \rm = 0^o$ (a), $\rm 70^o$ (b), and $\rm 90^o$ (c). Opposite to the bulge
(cf. Fig. 15), $A_{\lambda}$ increases with increasing $i$ at any $\lambda$
for the disk. Thus the absorption feature at $\rm 2175~\AA$ is washed out.
Note the abrupt change of $A_{\lambda}$ for $i \rm > 70^o$. The clumpiness of
the dusty ISM plays no role for $i \rm \ge 70^o$.\label{fig17}}
\end{figure}

\clearpage

\begin{figure}
\epsscale{.50}
\plotone{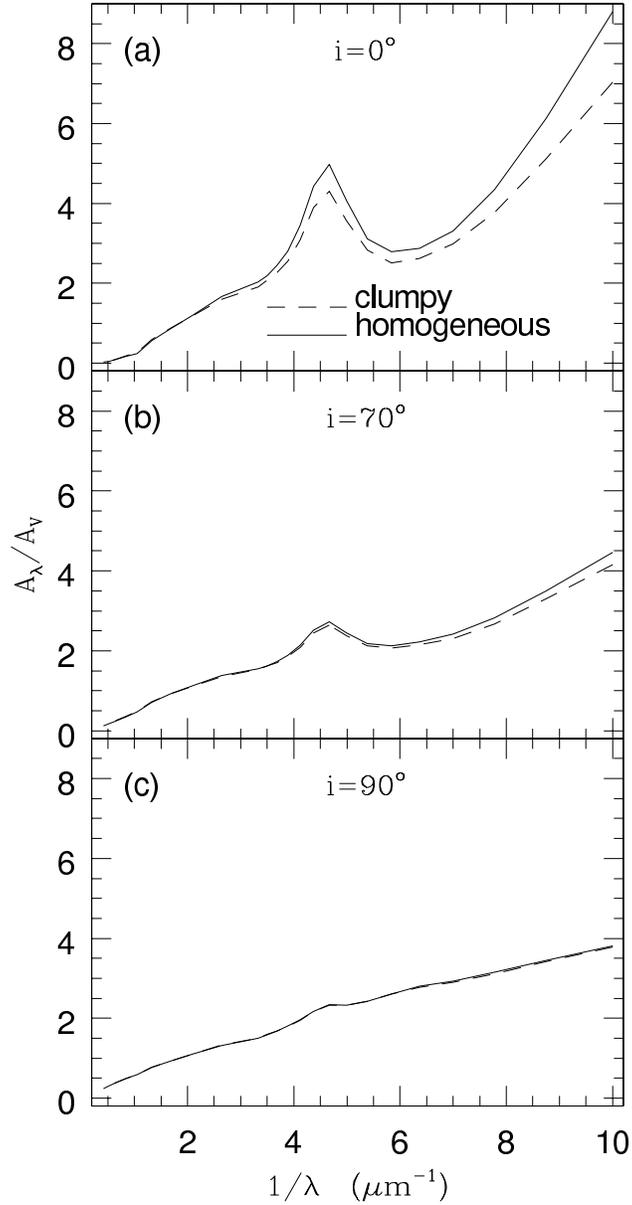}
\caption{$A_{\lambda}/A_V$ for the disk of a galaxy
with $\tau_V = 1.0$, as a function of the inclination
and the structure of the dusty ISM. Three cases are reproduced:
$i \rm = 0^o$ (a), $\rm 70^o$ (b), and $\rm 90^o$ (c). It emerges that
the relative increase of $A_{\lambda}$ is larger at progressively longer
wavelengths, when the disk is seen at increasingly higher inclinations.
\label{fig18}}
\end{figure}

\clearpage

\begin{figure}
\epsscale{.80}
\plotone{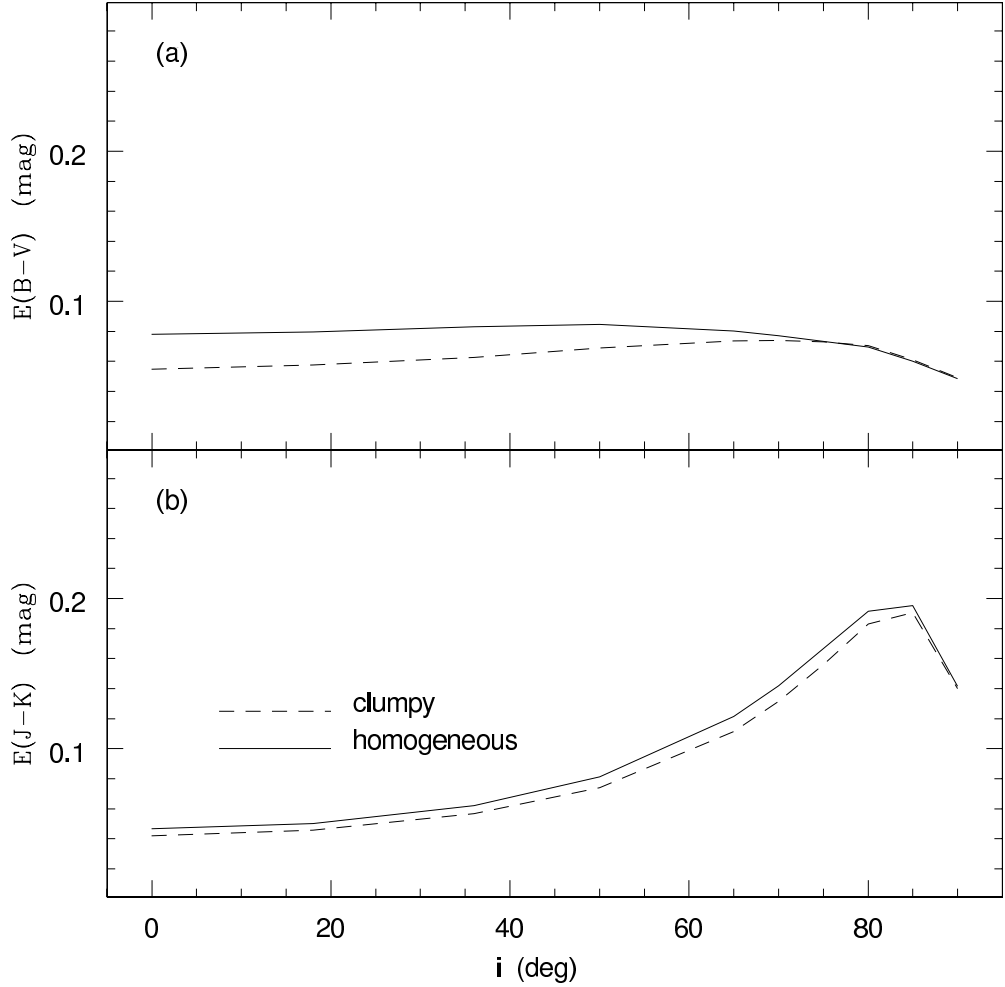}
\caption{$E(B-V)$ (a) and $E(J-K)$ (b) vs. $i$ for the bulge of
a late-type galaxy with $\tau_V = 1.0$, as a function of the structure of
the dusty ISM of the disk. In this case $E(B-V)$ is almost constant,
when $i$ increases up to $\rm \sim 60^o$, and decreases
for higher inclinations. Conversely, $E(J-K)$ strongly increases,
when $i$ increases up to $\rm \sim 80^o$, and decreases
for $i \rm > \sim 80^o$. This behavior may be understood when considering
that the behavior of the attenuation function with $i$ is different
in the optical and near-IR (cf. Fig. 15), and that the projection of the bulge
on the dusty disk becomes minimal when the galaxy is seen edge on.
\label{fig19}}
\end{figure}

\clearpage

\begin{figure}
\epsscale{.80}
\plotone{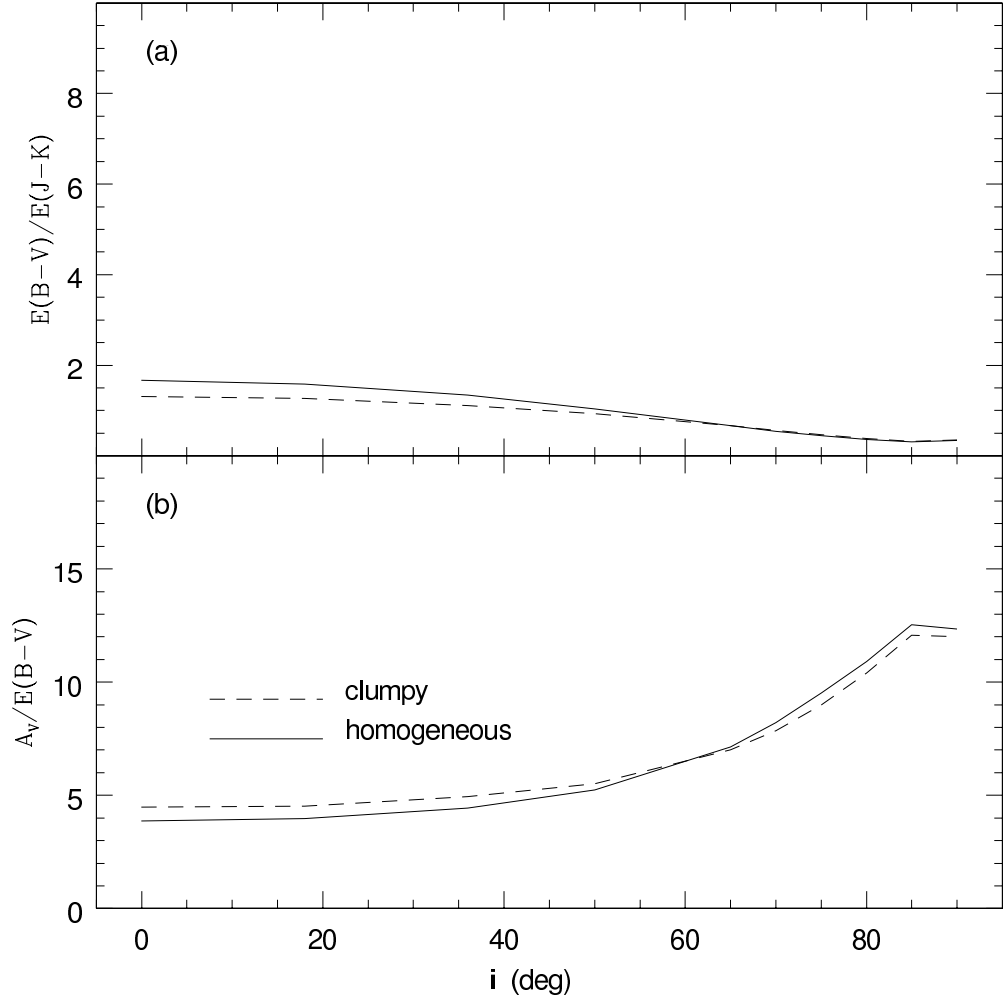}
\caption{$E(B-V)/E(J-K)$ (a) and $A_V/E(B-V)$ (b) vs. $i$
for the bulge of a late-type galaxy with $\tau_V = 1.0$, as a function of
the structure of the dusty ISM of the disk. The behavior of $E(B-V)/E(J-K)$
and $A_V/E(B-V)$ may be easily understood from Fig. 15.\label{fig20}}
\end{figure}

\clearpage

\begin{figure}
\epsscale{.80}
\plotone{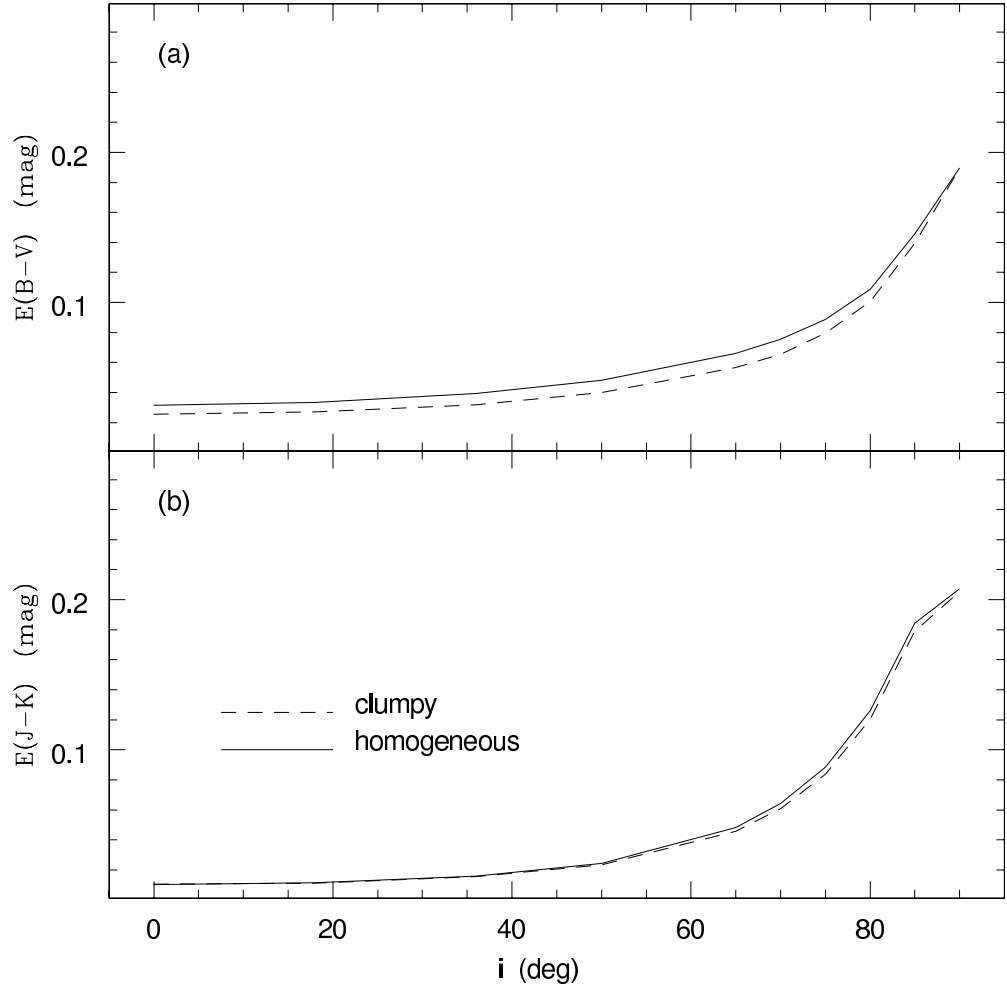}
\caption{$E(B-V)$ (a) and $E(J-K)$ (b) vs. $i$ for the disk of
a galaxy with $\tau_V = 1.0$, as a function of the structure of the dusty ISM.
Both the color excesses increase strongly and non-linearly
for $i \rm > 50^o$, as can be understood from Fig. 17.\label{fig21}}
\end{figure}

\clearpage

\begin{figure}
\epsscale{.80}
\plotone{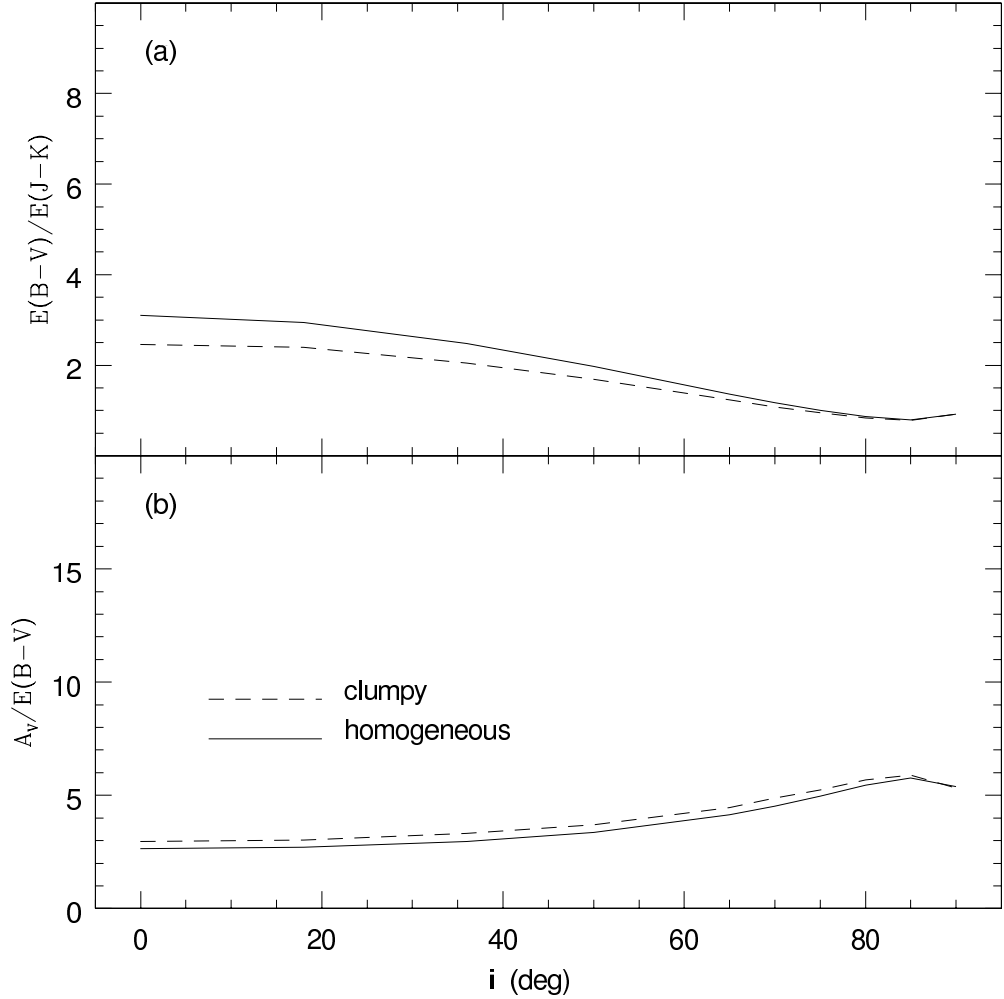}
\caption{$E(B-V)/E(J-K)$ (a) and $A_V/E(B-V)$ (b) vs. $i$
for the disk of a galaxy with $\tau_V = 1.0$, as a function of
the structure of the dusty ISM. The behavior of $E(B-V)/E(J-K)$
and $A_V/E(B-V)$ may be easily understood from Fig. 17. Note the small range of
$A_V/E(B-V)$ for the disk in comparison with the large one for the bulge
(Fig. 20).\label{fig22}}
\end{figure}

\clearpage

\begin{figure}
\epsscale{.50}
\plotone{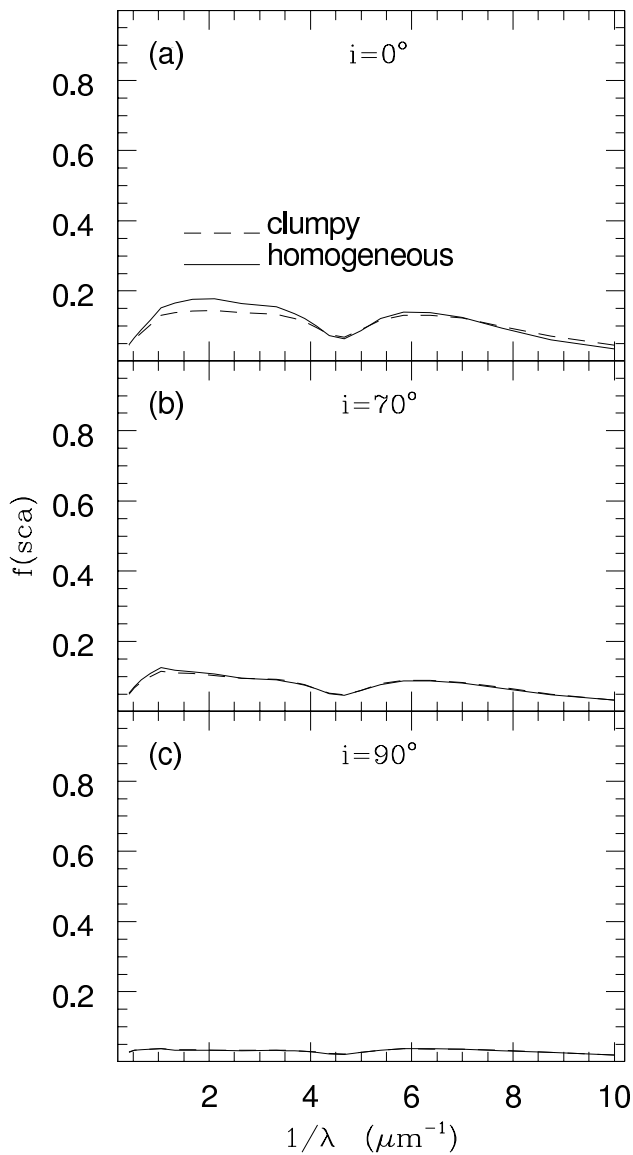}
\caption{The fraction of light which is emitted at a fixed $\lambda$
and is scattered towards the outside observer, $f(sca)$, for the bulge
of a late-type galaxy with $\tau_V = 1.0$, as a function of
the inclination of the galaxy, and the structure of the dusty ISM of the disk.
Three cases are reproduced: $i \rm = 0^o$ (a), $\rm 70^o$ (b),
and $\rm 90^o$ (c). $f(sca)$ decreases towards higher inclinations
not only because the escape probability of the scattered light decreases
(especially in the optical and UV), but also because a lower fraction of
the light traveling from the bulge towards the outside observer
is heavily affected by the dust in the disk.\label{fig23}}
\end{figure}

\clearpage

\begin{figure}
\epsscale{.50}
\plotone{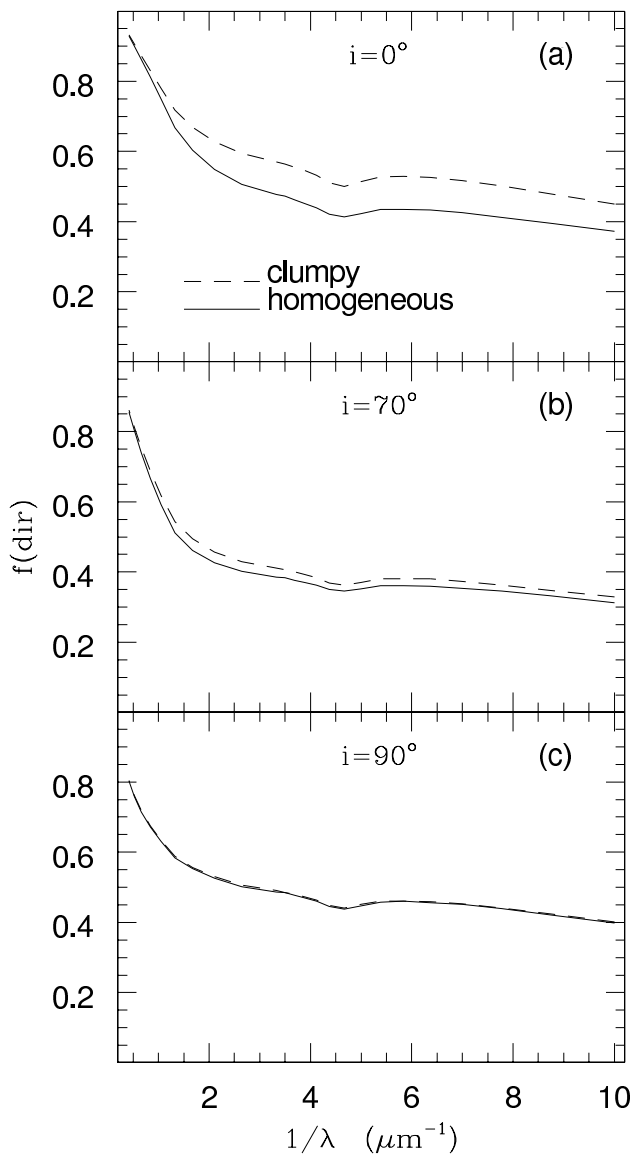}
\caption{The fraction of light which is emitted at a fixed $\lambda$
and is directly transmitted towards the outside observer, $f(dir)$,
for the bulge of a late-type galaxy with $\tau_V = 1.0$, as a function of
the inclination of the galaxy, and the structure of the dusty ISM of the disk.
Three cases are reproduced: $i \rm = 0^o$ (a), $\rm 70^o$ (b),
and $\rm 90^o$ (c). In general, $f(dir)$ decreases with increasing values
of the inclination. This trend is reversed when $i$ approaches $\rm 90^o$,
since the projection of the bulge on the dust disk becomes minimal
when the galaxy is seen edge on.\label{fig24}}
\end{figure}

\clearpage

\begin{figure}
\epsscale{.50}
\plotone{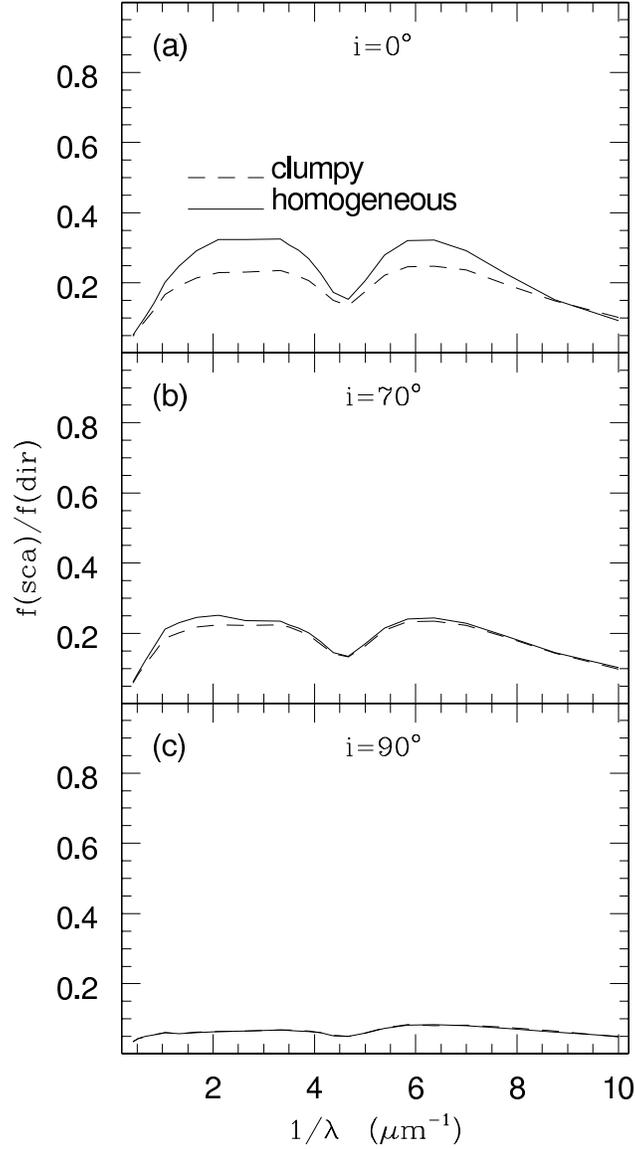}
\caption{$f(sca)/f(dir)$ for the bulge of a galaxy with $\tau_V = 1.0$,
as a function of the inclination of the galaxy and the structure
of the dusty ISM of the disk. Three cases are reproduced: $i \rm = 0^o$ (a),
$\rm 70^o$ (b), and $\rm 90^o$ (c). This behavior can be easily understood
from Fig. 23 and 24.\label{fig25}}
\end{figure}

\clearpage

\begin{figure}
\epsscale{.50}
\plotone{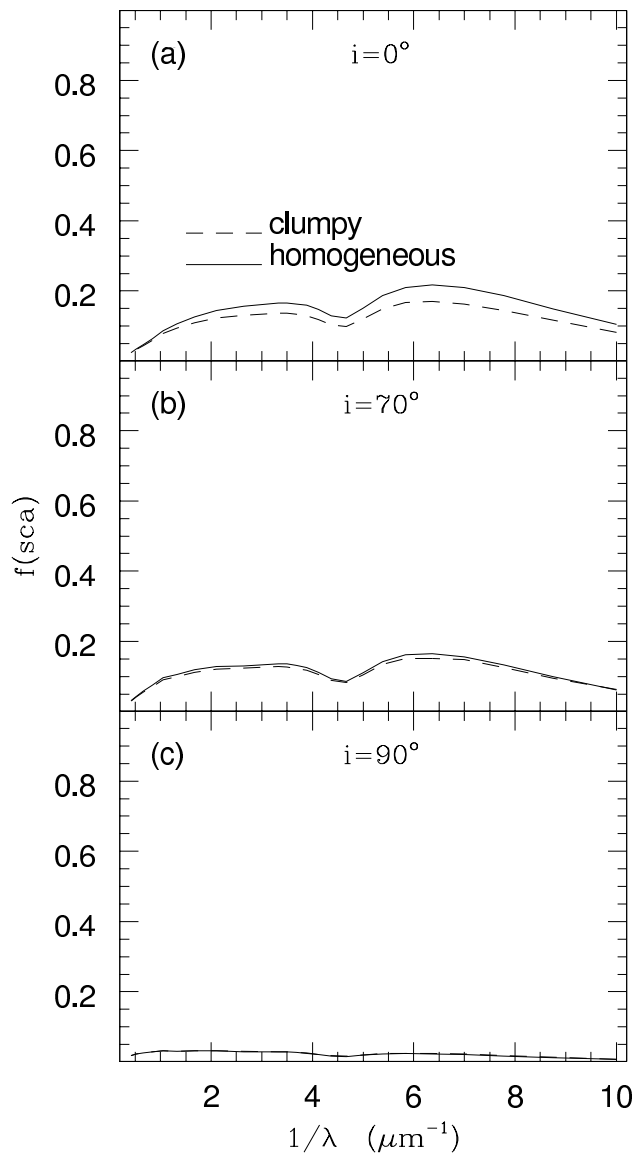}
\caption{$f(sca)$ for the disk of a galaxy with $\tau_V = 1.0$,
as a function of the inclination and the structure of the dusty ISM.
Three cases are reproduced: $i \rm = 0^o$ (a), $\rm 70^o$ (b), and $\rm 90^o$
(c). $f(sca)$ decreases with increasing $i$ owing to the increasing
blocking action of the dust along the line-of-sight (especially for
the far-UV photons), and the escape of the optical/near-IR photons
along lines of sight different from the observer's one.\label{fig26}}
\end{figure}

\clearpage

\begin{figure}
\epsscale{.50}
\plotone{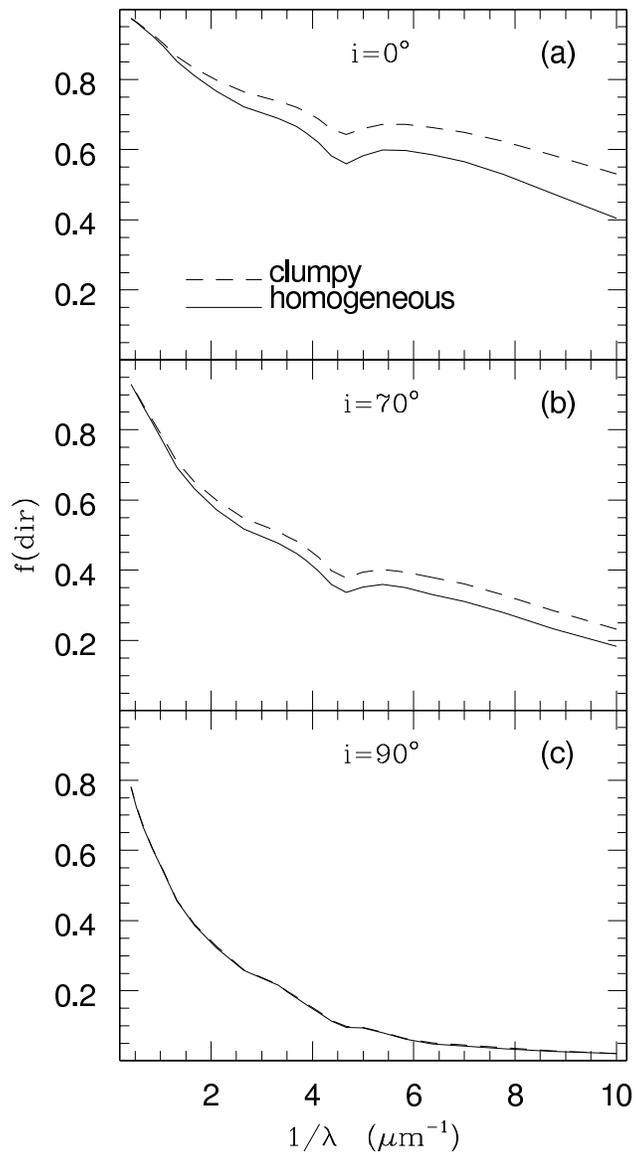}
\caption{$f(dir)$ for the disk of a galaxy with $\tau_V = 1.0$,
as a function of the inclination and the structure of the dusty ISM.
Three cases are reproduced: $i \rm = 0^o$ (a), $\rm 70^o$ (b), and $\rm 90^o$
(c). $f(dir)$ decreases with increasing $i$ owing to the increasing
blocking action of the dust along the line-of-sight (especially for
the far-UV photons), and the increasing importance of scattering
for the optical/near-IR photons.\label{fig27}}
\end{figure}

\clearpage

\begin{figure}
\epsscale{.50}
\plotone{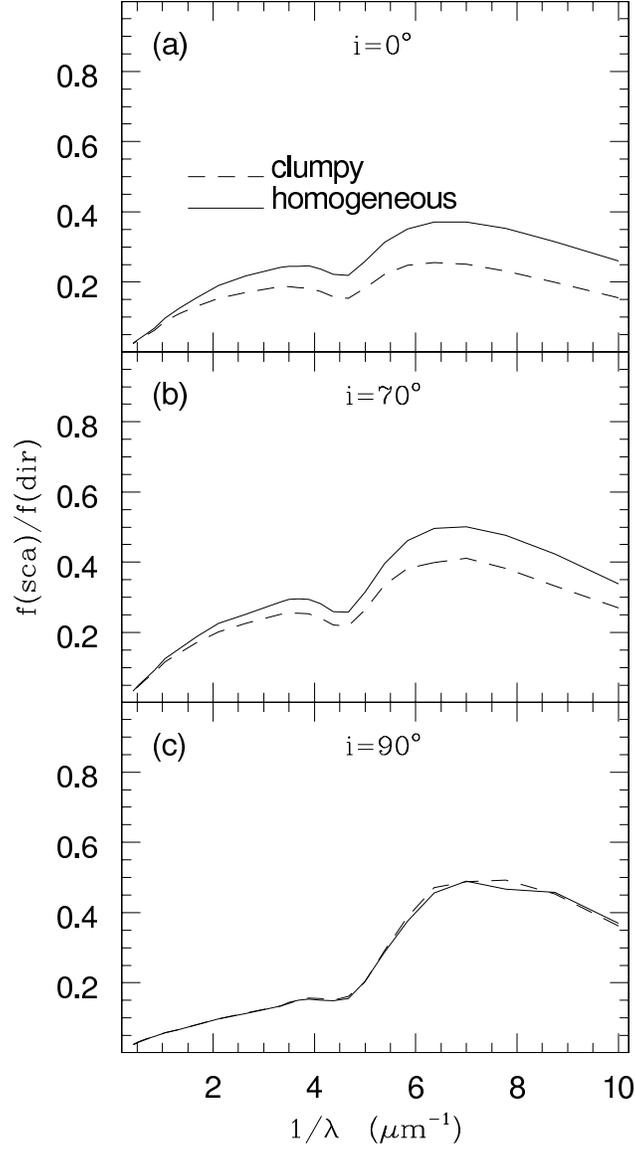}
\caption{$f(sca)/f(dir)$ for the disk of a galaxy with $\tau_V = 1.0$,
as a function of the inclination and the structure of the dusty ISM.
Three cases are reproduced: $i \rm = 0^o$ (a), $\rm 70^o$ (b), and $\rm 90^o$
(c). For the disk, $f(sca)/f(dir)$ increases slowly with increasing $i$,
except when $i$ approaches $\rm \sim 90^o$, opposite to the bulge
(cf. Fig. 25). For the edge-on disk, $f(sca)/f(dir)$ is reduced with respect to
e.g. the case of $i \rm = 70^o$, owing to the increased blocking action
of the dust along the line-of-sight (especially for the far-UV photons),
and the escape of the optical/near-IR photons along lines of sight
different from the observer's one.\label{fig28}}
\end{figure}

\clearpage

\begin{deluxetable}{rcccr}
\tabletypesize{\scriptsize}
\tablecaption{Main parameters for the extinction curve and the stellar disk}
\tablewidth{0pt}
\tablehead{
\colhead{$\lambda$} & \colhead{albedo} & \colhead{phase function asymmetry} &
\colhead{$\tau_{\lambda}$/$\tau_V$} & \colhead{scale height} \\ 
\colhead{[\AA]} & \colhead{} & \colhead{} & \colhead{} & \colhead{[pc]}
} 
\startdata
1000.0~ & 0.320~ & 0.800 & 5.328~ & 60.0~~~~~ \\
1142.9~ & 0.409~ & 0.783 & 3.918~ & 60.0~~~~~ \\
1285.7~ & 0.481~ & 0.767 & 3.182~ & 65.0~~~~~ \\
1428.6~ & 0.526~ & 0.756 & 2.780~ & 70.0~~~~~ \\
1571.4~ & 0.542~ & 0.745 & 2.584~ & 75.0~~~~~ \\
1714.3~ & 0.536~ & 0.736 & 2.509~ & 90.0~~~~~ \\
1857.1~ & 0.503~ & 0.727 & 2.561~ & 110.0~~~~~ \\
2000.0~ & 0.432~ & 0.720 & 2.843~ & 130.0~~~~~ \\
2142.9~ & 0.371~ & 0.712 & 3.190~ & 140.0~~~~~ \\
2285.7~ & 0.389~ & 0.707 & 2.910~ & 150.0~~~~~ \\
2428.6~ & 0.437~ & 0.702 & 2.472~ & 160.0~~~~~ \\
2571.4~ & 0.470~ & 0.697 & 2.194~ & 170.0~~~~~ \\
2714.3~ & 0.486~ & 0.691 & 2.022~ & 180.0~~~~~ \\
2857.1~ & 0.499~ & 0.685 & 1.905~ & 190.0~~~~~ \\
3000.0~ & 0.506~ & 0.678 & 1.818~ & 200.0~~~~~ \\
3776.8~ & 0.498~ & 0.646 & 1.527~ & 220.0~~~~~ \\
4754.7~ & 0.502~ & 0.624 & 1.199~ & 250.0~~~~~ \\
5985.8~ & 0.491~ & 0.597 & 0.909~ & 275.0~~~~~ \\
7535.8~ & 0.481~ & 0.563 & 0.667~ & 300.0~~~~~ \\
9487.0~ & 0.500~ & 0.545 & 0.440~ & 330.0~~~~~ \\
11943.5~ & 0.473~ & 0.533 & 0.304~ & 340.0~~~~~ \\
15036.0~ & 0.457~ & 0.511 & 0.210~ & 350.0~~~~~ \\
18929.2~ & 0.448~ & 0.480 & 0.145~ & 360.0~~~~~ \\
23830.6~ & 0.424~ & 0.445 & 0.100~ & 370.0~~~~~ \\
30001.0~ & 0.400~ & 0.420 & 0.069~ & 375.0~~~~~ \\
\enddata

\end{deluxetable}

\end{document}